\documentclass[twocolumn,superscriptaddress,amsmath,amssymb,aps,pra]{revtex4-1}

\usepackage{graphicx}
\usepackage{dcolumn}
\usepackage{bm}

\usepackage[colorlinks,urlcolor=blue,citecolor=blue,linkcolor=blue]{hyperref}

\usepackage{physics}
\usepackage{comment}
\usepackage{color}

\newcommand{\up}{\uparrow}
\newcommand{\down}{\downarrow}
\renewcommand{\k}{{\bf k}}

\newcommand{\p}{{\bf p}}

\newcommand{\q}{{\bf q}}

\newcommand{\eq}{\epsilon_{\q}}
\newcommand{\ek}{\epsilon_{\k}}
\newcommand{\ep}{\epsilon_{\p}}

\newcommand{\nn}{\nonumber}
\newcommand{\beq}{\begin{equation}}
\newcommand{\eeq}{\end{equation}}

\newcommand{\ri}{{\rm int}}
\newcommand{\rf}{\text{rf}}
\newcommand{\ks}{{\k\sigma}}
\newcommand{\e}{{\text{ej}}}
\newcommand{\ii}{{\text{inj}}}

\newcommand{\imp}{\text{imp}}
\newcommand{\OR}{\Omega_\text{R}}
\newcommand{\hH}{\hat{H}}
\newcommand{\hc}{\hat{c}}
\newcommand{\ha}{\hat{f}}
\newcommand{\hU}{\hat{U}}
\newcommand{\nB}{n_\text{B}}
\newcommand{\kB}{{k_\text{B}}}
\newcommand{\hr}{\hat{\rho}}

\newcommand{\m}{{\rm med}}
\newcommand{\ds}{\displaystyle}

\newcommand{\hd}{\hat{d}}

\newcommand{\hbc}{\hat{\bf c}}
\newcommand{\hep}{\hat{\varepsilon}}
\newcommand{\hp}{\hat{\phi}}
\newcommand{\hO}{\hat{O}}

\newcommand{\de}{{\rm Det}}
\newcommand{\hn}{\hat{n}}
\newcommand{\hsh}{\hat{h}}
\newcommand{\veF}{{E_F}}
\newcommand{\hN}{{\hat{N}}}

\newcommand{\hK}{\hat{K}}

\newcommand{\eb}{\epsilon_b}
\newcommand{\kb}{\kappa_b}

\newcommand{\en}{\epsilon}

\newcommand{\weizhe}[1]{{\color{black}#1}}

\begin{document}

\title{Theory of radio-frequency spectroscopy of impurities in quantum gases}

\author{Weizhe Edward Liu}
\affiliation{School of Physics and Astronomy, Monash University, Victoria 3800, Australia}
\affiliation{ARC Centre of Excellence in Future Low-Energy Electronics Technologies, Monash University, Victoria 3800, Australia}
\author{Zhe-Yu Shi}
\affiliation{School of Physics and Astronomy, Monash University, Victoria 3800, Australia}
\affiliation{State Key Laboratory of Precision Spectroscopy, East China Normal University, Shanghai 200062, China}
\author{Meera M.~Parish}
\affiliation{School of Physics and Astronomy, Monash University, Victoria 3800, Australia}
\affiliation{ARC Centre of Excellence in Future Low-Energy Electronics Technologies, Monash University, Victoria 3800, Australia}
\author{Jesper Levinsen}%
\affiliation{School of Physics and Astronomy, Monash University, Victoria 3800, Australia}
\affiliation{ARC Centre of Excellence in Future Low-Energy Electronics Technologies, Monash University, Victoria 3800, Australia}

\date{\today}

\begin{abstract}
We present a theory of radio-frequency spectroscopy of impurities interacting with a quantum gas at finite temperature. By working in the canonical ensemble of a single impurity, we show that the impurity spectral response is directly connected to the finite-temperature equation of state (free energy) of the impurity. We consider two different response protocols: ``injection’’, where the impurity is introduced into the medium from an initially non-interacting state; and ``ejection’’, where the impurity is ejected from an initially interacting state with the medium. We show that there is a simple mapping between injection and ejection spectra, which is connected to the detailed balance condition in thermal equilibrium. To illustrate the power of our approach, we specialize to the case of the Fermi polaron, corresponding to an impurity atom that is immersed in a non-interacting Fermi gas. For a mobile impurity with a mass equal to the fermion mass, we employ a finite-temperature variational approach to obtain the impurity spectral response. We find a striking non-monotonic dependence on temperature in the impurity free energy, the contact, and the radio-frequency spectra. For the case of an infinitely heavy Fermi polaron, we derive exact results for the finite-temperature free energy, thus generalizing Fumi's theorem to arbitrary temperature. We also determine the exact dynamics of the contact after a quench of the impurity-fermion interactions. Finally, we show that the injection and ejection spectra obtained from the variational approach compare well with the exact spectra, thus demonstrating the accuracy of our approximation method.
\end{abstract}

\maketitle

\section{Introduction}

\begin{figure}[hbt]
    \centering
    \includegraphics[width=0.95\columnwidth]{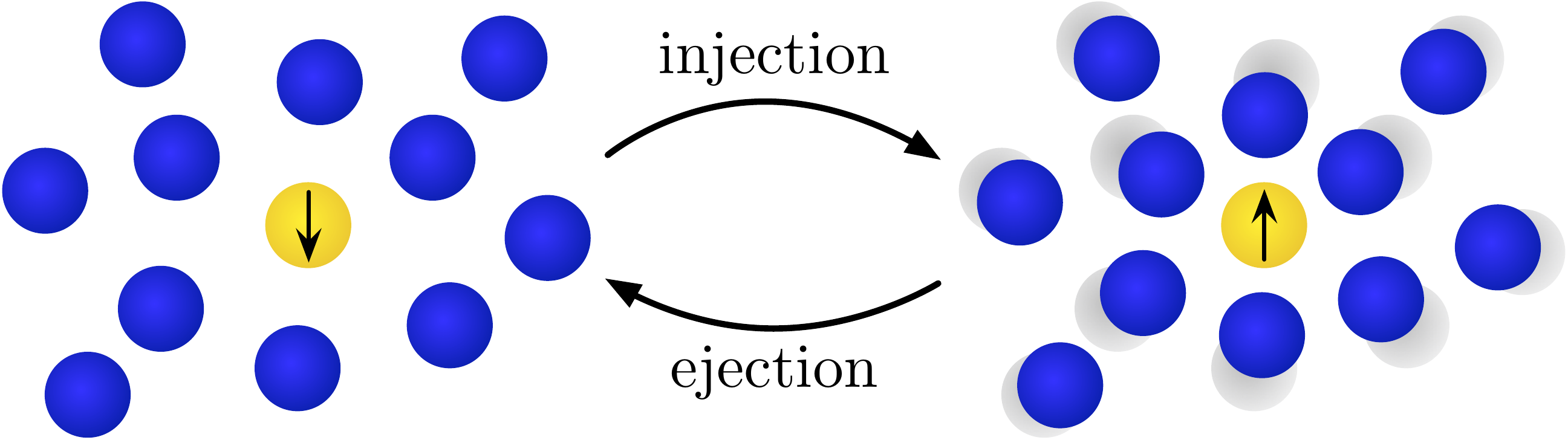}
    \caption{Schematic illustration of ejection and injection spectroscopy. A rf pulse drives a transition between the interacting ($\up$) and non-interacting ($\down$) states, and the transfer rate of atoms is measured as a function of the rf frequency.}
    \label{fig:illu}
\end{figure}

Radio-frequency (rf) spectroscopy is a powerful probe of many-body physics in trapped ultracold atomic gases. Most notably, it has been used to extract the pairing gap in a strongly interacting Fermi superfluid~\cite{Chin2004,Stewart2008} and to measure thermodynamic quantities such as the Tan contact~\cite{Sagi2012,Mukherjee2019}. There are typically two distinct rf spectroscopy schemes that can be experimentally deployed, as illustrated in Fig.~\ref{fig:illu}. The standard scheme~\cite{Gupta2003,Torma2014}, also known as direct or ejection spectroscopy, consists of preparing the interacting many-body system in equilibrium and applying an rf pulse to drive atoms from one spin state ($\up$) into another unoccupied spin state ($\down$) that does not interact with the surrounding atoms. Here the relevant observable is the fraction of the transferred atoms, which can be viewed as having been ejected from the interacting system. The opposite scheme~\cite{Frohlich2011,Cheuk2012} is known as inverse or injection spectroscopy in the literature. In this case, one starts with a non-interacting system and the observable is then the fraction of atoms excited from the non-interacting to interacting spin states.

Injection spectroscopy has provided a particularly important tool for studying the behavior of quasiparticles or \textit{polarons}~\cite{Massignan2014}, which are created by immersing impurity atoms in a quantum medium such as a degenerate Fermi~\cite{Schirotzek2009,Nascimbene2009,Kohstall2012,Koschorreck2012,Zhang2012,Cetina2015,Cetina2016,Scazza2017,Yan2019,Oppong2019} or Bose~\cite{Catani2012,Hu2016,Jorgensen2016,Camargo2018,Yan2019nature} gas. Here one can experimentally access impurity properties such as the quasiparticle energy and lifetime, 
as well as their dependence on system parameters such as temperature and impurity mass. In parallel with experimental progress, a variety of theoretical approaches have been used to calculate the impurity injection spectrum, including variational approaches~\cite{Li2014,Parish2016,Jorgensen2016,Shchadilova2016,Liu2019,Mistakidis2019prl,Field2020,Dzsotjan2019u}, impurity $T$-matrix methods~\cite{Massignan2008,Rath2013,Guenther2018,Hu2018}, the functional renormalization group~\cite{Schmidt2011}, the high-temperature virial expansion~\cite{Sun2017}, and diagrammatic quantum Monte Carlo (QMC)~\cite{Goulko2016}. However, there are comparatively few theoretical treatments of the impurity \textit{ejection} spectrum, which contains information about thermodynamic quantities such as the contact~\cite{Braaten2010,Yan2019}. Thus far, much of the work has focussed on the Fermi polaron at zero temperature~\cite{Punk2007,Veillette2008,Schneider2010,Schmidt2011}, and there have only recently been studies of the finite-temperature case using different $T$-matrix approximations~\cite{Tajima2019,Mulkerin2019}.

In this paper, which accompanies Ref.~\cite{Liu2020_short}, we show that the calculation of ejection spectra is considerably simplified by working in the canonical ensemble, where one has a single impurity in the medium. Specifically, we find that the impurity ejection spectrum is related to the corresponding injection spectrum via a simple factor involving the finite-temperature equation of state (free energy) of the impurity. This is not obvious at first sight since the two impurity spectra can have qualitatively different features; for example, the ejection spectrum exhibits a high-frequency tail that is related to the contact,  whereas this feature appears to be absent in the injection spectrum. Moreover, this mapping holds regardless of the medium type or the system dimensionality.

As a concrete example, we apply our spectral relation to the well-studied case of the Fermi polaron. Here, for an infinitely heavy impurity, we can compute both injection and ejection spectra exactly~\cite{Goold2011,Knap2012,Schmidt2018}, which allows us to confirm our prediction as well as to benchmark approximation methods. We also calculate the impurity free energy and contact exactly for arbitrary temperature, thus generalizing Fumi's theorem~\cite{Fumi1955} to finite temperature. To gain further insight into the exact solution, we determine how the contact relaxes towards its equilibrium value after a quench of the impurity-fermion interactions to unitarity. We show that the relaxation dynamics are fast, i.e., comparable to the Fermi time scale, which is consistent with experimental observations of the contact dynamics in a unitary Bose gas~\cite{Fletcher2017}.

For the case of a mobile impurity with mass the same as the fermion mass, we obtain approximate injection and ejection spectra using a recently developed finite-temperature variational approach~\cite{Liu2019}. Focussing on the quasiparticle peak in the spectrum, we extract the polaron energy as a function of temperature, and we compare it with the infinite-mass case. We also compute the impurity free energy and contact using our injection-ejection relation and the sum rules for the spectral functions. We find that both the free energy and the contact display a striking non-monotonic dependence on temperature, which differs from the corresponding infinite-mass results but agrees with recent experiment~\cite{Yan2019,Liu2020_short}.

The paper is organised as follows. In Sec.~\ref{sec:impspectra}, we present the definitions of ejection and injection spectra and their properties, including the connection to the standard Green's function approach. In Sec.~\ref{sec:thermodyn}, we consider the thermodynamic quantities in the impurity problem, particularly the impurity free energy and contact. While Sections~\ref{sec:impspectra} and \ref{sec:thermodyn} are applicable to the general impurity problem, in Sec.~\ref{sec:Fermipolaron} we specialize to the case of the mobile Fermi polaron, for which we derive injection and ejection spectra as well as thermodynamic properties. In Sec.~\ref{sec:fixedimp}, we then consider the limit of an infinitely heavy impurity in a Fermi gas for which we obtain exact spectra, explicitly demonstrating the relation between the injection and ejection protocols. Additionally, we generalize Fumi's theorem to finite temperature, which allows us to obtain the impurity free energy as well as the contact for arbitrary interaction strength and temperature. We conclude in Sec.~\ref{sec:concl}.

\section{Impurity spectral functions}\label{sec:impspectra}

We first discuss the relationship between the impurity injection and ejection spectral protocols, illustrated in Fig.~\ref{fig:illu}. Both have been utilized in experiments on polarons: the impurity injection spectrum has been measured in Refs.~\cite{Kohstall2012,Cetina2016,Oppong2019} in the context of the Fermi polaron and in Refs.~\cite{Hu2016,Jorgensen2016,Camargo2018} for the Bose polaron, while ejection spectra have been measured in Refs.~\cite{Schirotzek2009,Koschorreck2012,Yan2019} and \cite{Yan2019nature} for the Fermi and Bose polarons, respectively. 

The injection spectrum can be calculated using a variety of theoretical methods (as discussed in the introduction), whereas the ejection spectrum is more challenging to compute since one must consider a finite density of impurities within the usual grand-canonical formulation~\cite{Schmidt2011}. By contrast, we use here the canonical ensemble for the impurity where we can naturally take the limit of a single impurity, while still employing the grand canonical ensemble for the medium. We stress that this is not merely a technical detail. As we shall explicitly demonstrate, our formalism is simpler than previous approaches, and allows us to arrive at an important relationship between the different spectroscopic protocols.

We emphasize that the formalism introduced in this section is very general, as it applies to a single impurity immersed in any medium in thermal equilibrium, independent of dimensionality. That is, the medium can be fermionic, bosonic, two-component fermions in the BCS-BEC crossover below or above the critical temperature for superfluidity, or even systems beyond cold atoms. Furthermore, we also expect our results to hold for a finite density of impurities, as long as these are uncorrelated.

\subsection{Definitions of spectral functions}

Although the experimental protocols of injection and ejection spectroscopy appear different, we can still introduce a single Hamiltonian (with different initial states) to describe these:
\begin{align}
    \hH(t)
    &=\hH_\ri+\hH_\rf(t)\theta(t).\label{hamiltonian}
\end{align}
Here
\begin{align}
    \hH_\rf(t)&=\OR\sum_\k\left[\hc_{\k\down}^\dagger\hc_{\k\up}e^{-i\omega t}
    +\hc_{\k\up}^\dagger\hc_{\k\down}e^{i\omega t} \right],
\end{align}
describes the Rabi coupling due to an external rf field~\footnote{In typical cold atom experiments, the relevant transition frequencies lie in the radio frequency range, and we will therefore use the terminology relevant to radio frequency spectroscopy throughout this work.} that is turned on at time $t=0$ ($\theta$ is the unit step function). Here, $\OR$ is the strength of the Rabi coupling, $\omega$ is the frequency measured from the bare $\down$--$\up$ transition, and we have applied the rotating wave approximation. In the impurity creation operator $\hc^\dag_{\k\sigma}$, $\k$ is the momentum while $\sigma=\down,\up$ represents the impurity (pseudo) spin degree of freedom. The associated dispersion is $\epsilon_\k = k^2/(2M)$ where $k\equiv |\k|$, $M$ is the impurity mass, and we work in units where both $\hbar$ and $\kB$ are 1. We write the remainder of the Hamiltonian as
\begin{align}
    \hH_\ri&= \underbrace{\hH_\m + \hH_\imp}_{\hH_0} +\hU,
\end{align}
where
\begin{align}
\hH_\imp=\sum_\ks \epsilon_\k \hc_\ks^\dagger \hc_\ks
\end{align}
is the non-interacting impurity Hamiltonian. Since at this stage we are discussing properties of the impurity spectral response that are independent of the precise nature of the medium, we do not explicitly specify the medium-only Hamiltonian $\hH_\m$ or the impurity-medium interaction Hamiltonian $\hU$. Throughout this paper, we will assume that only the impurity in spin state $\sigma=\up$ interacts with the medium.

In the following, we consider a single impurity immersed in a medium containing a macroscopic number of particles. For $t<0$, the impurity is in a well-defined spin state and the system is in thermal equilibrium at a temperature $T$. As illustrated in Fig.~\ref{fig:illu}, once the rf field is applied at $t>0$, ejection spectroscopy measures the particle transfer rate ${\cal I}_\e(\omega)$ starting from the interacting state $\up$ to the non-interacting state $\down$. Conversely, injection spectroscopy measures the transition rate ${\cal I}_\ii(\omega)$ from the non-interacting state $\down$ to the interacting state~$\up$.

Within linear response~\cite{Fetter}, we calculate the transfer rates using Fermi's golden rule. This leads to
\begin{subequations}
    \label{eq:I}
\begin{align}
    {\cal I}_\e(\omega)&\equiv 2\pi\OR^2 I_\e(\omega)= 2\pi\OR^2\sum_\p A_\e(\p,\omega),\\[2ex]
    {\cal I}_\ii(\omega)&\equiv 2\pi\OR^2 I_\ii(\omega)= 2\pi\OR^2\sum_\p \nB(\p)A_\ii(\p,\omega),
\end{align}
\end{subequations}
where we introduce the ejection and injection impurity spectral functions $I_\e$, $A_\e$ and $I_\ii$, $A_\ii$. The sums are over impurity momentum $\p$, and 
\begin{align}
\label{eq:nB}
 \nB(\p)=e^{-\beta\epsilon_\p}/Z_\imp   
\end{align}
is the Boltzmann momentum distribution density of the non-interacting impurity, with $Z_\imp=\sum_\p e^{-\beta\epsilon_\p}$ the single impurity partition function~\cite{foot:volume}. Here we have defined $\beta\equiv 1/T$. Most experiments performing rf spectroscopy on impurities have measured the momentum averaged spectra $I_\e$ and $I_\ii$, although the momentum-resolved ejection spectrum $A_\e$ of Fermi polarons in two dimensions was measured in Ref.~\cite{Koschorreck2012} using angle-resolved photo-emission spectroscopy, while the injection spectrum $A_\ii$ at zero momentum was measured for the three-dimensional Bose polaron in Ref.~\cite{Jorgensen2016} where impurities were transferred from a Bose-Einstein condensate.

The spectral functions introduced in Eq.~\eqref{eq:I} take the form
\begin{subequations}
\label{eq:fermigrule}
\begin{align}
    \label{eq:fermigrule_a}
    A_\e(\p,\omega)&
    \!=\sum_{n,\nu}\frac{e^{-\beta E_{\nu}}}{Z_\ri}|\!\bra{n}\hc_{\p\up}\ket{\nu}\!|^2 \delta(E_{\nu n\p}+\omega),\\
    \label{eq:fermigrule_b}
    A_\ii(\p,\omega)&
    \!=\sum_{n,\nu}\frac{e^{-\beta E_n}}{Z_\m}|
    \!\bra{n}\hc_{\p\up}\ket{\nu}\!|^2
    \delta(E_{\nu n\p}-\omega).
\end{align}
\end{subequations}
Here, $E_\nu$ ($|\nu\rangle$) corresponds to the eigen energy (state) of the interacting Hamiltonian $\hH_\ri$ with a single impurity in state $\sigma=\up$, $E_n$ ($|n\rangle$) corresponds to the eigen energy (state) for the noninteracting Hamiltonian of the medium without an impurity (i.e., the impurity vacuum), and $Z_\ri\equiv\sum_\nu e^{-\beta E_\nu}$ and $Z_\m\equiv \sum_n e^{-\beta E_n}$ are the corresponding partition functions. To distinguish the possible eigenstates, we represent the interacting states with the Greek letter $\nu$ and the non-interacting medium-only states with the Latin letter $n$. The delta functions ensure energy conservation where we have defined $E_{\nu n\p} \equiv E_\nu - E_n -\epsilon_\p$. Likewise, the matrix element $\bra{n}\hc_{\p\up}\ket{\nu}$ vanishes unless momentum is conserved.

In the following, it is useful to introduce the density matrix associated with the interacting impurity and medium system
\begin{align}
    \hr_\ri=e^{-\beta \hH_\ri}/Z_\ri
\end{align}
as well as a density matrix of the medium only
\begin{align}
    \hr_\m=e^{-\beta \hH_\m}/Z_\m.
\end{align}
With these definitions, we then define the trace over interacting states, $\Tr[\hr_\ri\cdots]\equiv\sum_\nu \bra{\nu}\hr_\ri \cdots \ket{\nu}$, as well as the trace over medium-only states $\Tr[\hr_\m\cdots]\equiv\sum_n \bra{n} \hr_\m\cdots \ket{n}$.

\subsection{Properties of spectral functions}
\label{sec:specprops}

The fact that $A_\e$ and $A_\ii$ are related by a single Hamiltonian~\eqref{hamiltonian} suggests that there may exist an internal relation between the two spectral functions. Indeed, by utilizing the properties of the $\delta$-functions in Eq.~\eqref{eq:fermigrule}, we obtain the detailed balance condition,
\begin{align}
    A_\e(\p,\omega)=
    \frac{Z_\m Z_\imp}{Z_\ri}e^{\beta\omega} n_{\rm B}(\p)A_\ii(\p,-\omega).\label{eq:detailed_balance}
\end{align}
For details of this derivation, see App.~\ref{app:Aejinj_def}.
After summing up the impurity momentum contributions according to Eq.~\eqref{eq:I}, we find
\begin{equation}\label{eq:detailed_bal_total}
    I_\e(\omega) = \frac{Z_\m Z_\imp}{ Z_\ri} e^{\beta\omega} I_\ii (-\omega).
\end{equation}
We emphasize that Eqs.~\eqref{eq:detailed_balance} and \eqref{eq:detailed_bal_total} are key equations in this paper. They indicate that in the single-impurity limit, the two distinct experimental protocols of ejection and injection spectroscopy are closely related, such that we only need to know one to obtain the other. In fact, the prefactor in both relations, $\frac{Z_\m Z_\imp}{ Z_\ri}$, is related to the impurity free energy and can be independently measured, as we discuss in Sec.~\ref{sec:thermodyn}. Importantly, $A_\ii$ is typically simpler to calculate because it involves a thermal average with respect to the medium-only Hamiltonian $\hH_\m$, rather than an average that involves all the interacting eigenstates. We also note that it is possible to arrive at a detailed balance condition for the case where both impurity spin states interact with the medium. This is discussed in App.~\ref{app:detailed_int}.

By integrating Eq.~\eqref{eq:fermigrule} over all frequencies, we immediately obtain the following sum rules,
\begin{subequations}
\label{sum rule}
\begin{align}
    \label{sum rule_a}
    \int d\omega A_\e(\p,\omega)& =n_\ri(\p),
    \\[3ex]
    \label{sum rule_b}
    \int d\omega A_\ii(\p,\omega)&=1,
\end{align}
\end{subequations}
where $n_\ri(\p)\equiv \Tr[\hr_\ri\hc^\dagger_{\p\up}\hc_{\p\up}]$ is the impurity momentum distribution density of the interacting many-body system. From the above sum rules, we see that~\cite{foot:volume}
\begin{align}
    A_\e(\p,\omega)\sim V^{-1},\qquad A_\ii(\p,\omega)\sim V^0,\label{order}
\end{align}
where $V$ is the system volume. Thus, in the thermodynamic limit, $A_\e (\p,\omega) \to 0$ while $A_\ii (\p,\omega)$ remains finite. This is consistent with the vanishing ejection spectrum that one would obtain from a grand canonical $T$-matrix approach when taking the single impurity limit, see, e.g., Ref.~\cite{Schneider2010}. It is also consistent with the detailed balance condition in Eq.~\eqref{eq:detailed_balance}: We have $Z_\imp\sim V$ (and hence $n_{\rm B}\sim V^{-1}$), and  furthermore $Z_\m /Z_\ri\sim V^{-1}$ since the interacting states have one more particle (the impurity) than the medium-only states. 

We can additionally sum over impurity momentum to find
\begin{align}
    \label{sum rule_c}
    \ds \int d \omega\sum_\p n_{\rm B}(\p) A_\ii(\p,\omega)&=1.
\end{align}
On the other hand, we can have situations where the spin-$\up$ impurity-medium interactions transfers the impurity to a different state, for instance a closed channel in a two-channel model. Only if this is \textit{not} the case do we have the corresponding sum rule $\ds \int d \omega\sum_\p A_\e(\p,\omega)=1.$ This point is discussed with a concrete example in Sec.~\ref{sec:Fermipolaron}.

\subsection{Relation to impurity Green's function}

We now relate our results to the impurity Green's function, which can be approximated using standard diagrammatic techniques. A series of straightforward manipulations (see App.~\ref{app:spectral} for details) allows us to arrive at an alternative representation for $A_\e$ and $A_\ii$ introduced in Eq.~\eqref{eq:fermigrule}:
\begin{subequations}\label{spectral_functions_all}
\begin{align}
    \label{eq:Aej2}
    A_\e(\p,\omega)& = \Re\int_0^\infty \frac{dt}\pi e^{i(\omega-\epsilon_\p) t}\Tr\left[\hr_\ri\hc_{\p\up}^\dagger(t)\hc_{\p\up}(0) \right],\\[3ex]
     \label{eq:Ainj2}
    A_\ii(\p,\omega)& =\Re\int_0^\infty \frac{dt}\pi e^{i(\omega+\epsilon_\p) t}\Tr\left[\hr_\m\hc_{\p\up}(t)\hc^\dagger_{\p\up}(0) \right],
\end{align}
\end{subequations}
where $\hc_{\p\up} (t)$ is the time-dependent impurity operator in the Heisenberg picture
\begin{align}
    \hc_{\p\up}(t)\equiv e^{i\hH_\ri t}\hc_{\p\up}e^{-i\hH_\ri t}.\label{eq:heisenberg_operator}
\end{align}
Therefore, impurity spectroscopy is closely connected to the time-evolution of the impurity operator.

We next define the retarded time-dependent impurity Green's function in the grand canonical ensemble for the medium, but restricted to a single impurity:
\begin{align}
    G^{\rm R}_\up(\p,t)
&\weizhe{=-i\,\theta(t)\Tr\left[\hr_\m\hc_{\p\up}(t)\hc^\dagger_{\p\up}(0) \right]}.
    \label{eq:Gt}
\end{align}
The spectral function is defined in terms of the Fourier transform
\begin{align}
    A(\p,\omega)=- \Im[G_\up^{\rm R}(\p,\omega)]/\pi=- \Im\int \frac{dt}\pi\, e^{i\omega t}G_\up^{\rm R}(\p,t).
    \label{eq:Aomega}
\end{align}
Note that throughout our discussions of Green's functions, we implicitly take $\omega$ to have an infinitesimal positive imaginary part~\cite{Fetter}. By comparing Eqs.~\eqref{eq:Gt} and \eqref{eq:Aomega} with Eq.~\eqref{eq:Ainj2}, we see that the impurity injection spectral function is simply related to the spectral function via
\begin{align}
    A_\ii (\p,\omega)=A(\p,\omega+\ep)=- \Im[G_\up^{\rm R}(\p,\omega+\ep)]/\pi.
\end{align}

\subsection{Finite density of impurities}
\label{sec:finitedens}

Detailed balance conditions are usually formulated using the grand canonical ensemble for both the impurity and the medium, see, e.g., Ref.~\cite{Haussmann2009}. This has the advantage that it explicitly allows one to treat a finite density of impurities; however it also complicates the calculations~\cite{Schmidt2011}. At a technical level, the finite impurity density results in additional poles in the complex energy plane. For completeness, we now outline the connection between these two approaches. 

In this subsection, we thus consider a finite density of impurities at chemical potential $\mu_\up$. The grand canonical spin-$\up$ retarded Green's function is then~\cite{Fetter}
\begin{align}
    {\cal G}^\text{R}_\up(\p,t)=-i\, \theta(t) \Tr[\hr_{\rm G} \{ e^{i\hK t}\hc_{\p\up} e^{-i\hK t},\hc^\dagger_{\p\up}\}\weizhe{_{\pm}}],
    \label{eq:Gfinitedens}
\end{align}
where $\hat{K}\equiv \hH_\ri - \mu_\up \hN_\up$ with $\hN_\up$ the spin-$\up$ number operator \weizhe{and $\{A,B\}_{\pm} = AB \pm BA$ with $+$ and $-$ for fermions and bosons, respectively}. The trace is over all Fock states (i.e., not restricted to either 0 or 1 impurity) which, to be concise, we still label by a Greek letter $\nu$, and $\hr_{\rm G}=e^{-\beta\hK}/{\cal Z}_{\rm G}$ with ${\cal Z}_{\rm G}=\sum_\nu e^{-\beta E_\nu}$.

Similarly to the single impurity scenario, Eq.~\eqref{eq:Aomega}, the spin-$\up$ spectral function is~\cite{Haussmann2009}
\begin{align}
    {\cal A}(\p,\omega) & =-\Im[{\cal G}^{\rm R}_\up(\p,\omega)]/\pi \nn \\ 
    &=- \Im\int \frac{dt}\pi\, e^{i(\omega-\mu_\up)t}{\cal G}_\up^{\rm R}(\p,t),
    \label{eq:Afinitedens}
\end{align}
where we find it convenient to measure the frequency from the chemical potential of spin $\up$ \weizhe{particles}. The spectral function can be decomposed into two terms
\begin{align}
    {\cal A}(\p,\omega) = {\cal A}_+ (\p,\omega) \weizhe{\,\pm\,} {\cal A}_- (\p,\omega).
\end{align}
${\cal A}_+$ and ${\cal A}_-$ are known as the particle and hole parts of the excitation spectrum, respectively. They take the form (for details, see App.~\ref{app:spectral_GC})
\begin{subequations}\label{eq:spectral_GC}
\begin{align}
    {\cal A}_+ (\p,\omega) & = \sum_{\xi,\nu} \abs{\bra{\xi} \hc_{\p\up} \ket{\nu}}^2 \delta (\omega + E_\xi - E_\nu ) \expval{\hr_{\rm G}}{\xi},\\
    {\cal A}_- (\p,\omega) & = \sum_{\xi,\nu} \abs{\bra{\xi} \hc_{\p\up} \ket{\nu}}^2 \delta (\omega + E_\xi - E_\nu ) \expval{\hr_{\rm G}}{\nu}.
\end{align}
\end{subequations}
As discussed in, e.g., Ref.~\cite{Haussmann2009}, the particle and hole parts of the spectrum are related by the detailed balance condition
\begin{align}
    {\cal A}_-(\p,\omega)=e^{-\beta (\omega-\mu_\up)}{\cal A}_+(\p,\omega),
    \label{eq:detailed_GC}
\end{align}
which follows from analyzing the Lehmann representation of the Green's function~\cite{Fetter} (see also App.~\ref{app:spectral_GC}). This equation is reminiscent of the momentum-resolved single-impurity detailed balance condition in Eq.~\eqref{eq:detailed_balance}, and can therefore be thought of as a grand canonical ensemble version of Eq.~\eqref{eq:detailed_balance}. 

Let us now discuss how we can relate the grand canonical ensemble results to the case of a single impurity. The key observation is that in this limit, the impurity chemical potential $\mu_\up\to-\infty$. \weizhe{This is because, in the grand canonical ensemble, any finite $\mu_\up$ leads to a finite impurity density.} Thus, from Eq.~\eqref{eq:detailed_GC} we see that ${\cal A}_-(\p,\omega)\to0$. More generally, we find that states with multiple impurities are suppressed in the traces in Eq.~\eqref{eq:spectral_GC} by positive powers of $e^{\beta\mu_\up}$. Therefore, in the limit of a single impurity, we simply have (App.~\ref{app:spectral_GC})
\begin{align}
    {\cal A}(\p,\omega)={\cal A}_+(\p,\omega)=A(\p,\omega),
\end{align}
and hence
\begin{subequations}
\label{eq:AGCvsAC}
\begin{align}
    \label{eq:Aplus_def}
    {\cal A}_+(\p,\omega) &= A_\ii (\p,\omega-\epsilon_\p), \\[1ex]
    \label{eq:Aminus_def}
    {\cal A}_-(\p,\omega) &= \left(\sum_\q e^{-\beta(\eq-\mu_\up)}\right)\frac{Z_\ri}{Z_\m Z_\imp}A_\e (\p,\epsilon_\p-\omega).
\end{align}
\end{subequations}
In the second of these equations we simply used the detailed balance conditions, Eq.~\eqref{eq:detailed_balance} and~\eqref{eq:detailed_GC}.

While Eq.~\eqref{eq:AGCvsAC} provides a direct relationship between the particle/hole parts of the impurity spectrum and the injection/ejection impurity spectral functions, we emphasize that the canonical ensemble formulation has several clear advantages. First, for a finite density of impurities, it is not possible to derive a detailed balance condition similar to Eq.~\eqref{eq:detailed_bal_total} for the momentum averaged spectral response. Since the majority of experimental setups are limited to momentum-averaged spectra, this is a major difference. Second, as we discuss in the following section, the single impurity limit also allows us to directly extract thermodynamic properties.

\section{Impurity thermodynamics}
\label{sec:thermodyn}

Before turning to a concrete example, we briefly discuss the thermodynamic properties of the medium with a single impurity, namely the free energy and the contact.

\subsection{Free energy}

We start by defining the impurity free energy as the difference between the free energy of the interacting and non-interacting systems: $\Delta F\equiv F-F_0$. The free energy is related to the partition functions defined in Sec.~\ref{sec:impspectra}, i.e., $Z_\ri\equiv e^{-\beta F}$ and $Z_\m Z_{\rm imp}\equiv e^{-\beta F_0}$. Therefore,
\begin{align}
    \Delta F=F-F_0=T\ln{\frac{Z_\m Z_{\rm imp}}{Z_\ri}}.
    \label{eq:FE}
\end{align}

The ratio of partition functions in Eq.~\eqref{eq:FE} also appears in the detailed balance conditions, Eqs.~\eqref{eq:detailed_balance} and \eqref{eq:detailed_bal_total}. We can therefore rewrite these as
\begin{subequations}
\label{eq:detailed_final}
\begin{align}
    A_\e(\p,\omega) &=
    e^{\beta\Delta F}e^{\beta\omega} n_{\rm B}(\p)A_\ii(\p,-\omega),
    \label{eq:detailed_ej_F}
    \\
    I_\e(\omega) & = e^{\beta\Delta F} e^{\beta\omega} I_\ii (-\omega).
    \label{eq:detailed_ej_I}
\end{align}
\end{subequations}
In particular, the detailed balance conditions indicate that the impurity free energy can be measured by simply comparing the injection and ejection spectra, since
\begin{align}
    \Delta F=-\omega+T\ln{\frac{I_\e(\omega)}{I_\ii(-\omega)}}
\end{align}
is independent of $\omega$. We expect that the free energy may be accurately measured in this manner close to a quasiparticle peak in the injection spectrum, as long as the energy difference between the peak and the $T=0$ ground state does not greatly exceed the temperature. We also note that the quasiparticle peaks generally approach zero frequency (i.e., the bare $\down$--$\up$ transition frequency) with increasing temperature, and therefore this provides an interesting reference point where we simply have
\begin{align}
    \Delta F=T\ln{\frac{I_\e(0)}{I_\ii(0)}}.
\end{align}

The sum rules allow us to derive two useful relations for the impurity free energy. From the detailed balance condition in Eq.~\eqref{eq:detailed_ej_F} and the sum rule in Eq.~\eqref{sum rule_a}, we find 
\begin{align}
    &\hspace{-10mm}e^{-\beta\Delta F}Z_\imp\sum_\p  \Tr[\hr_\ri\hc^\dagger_{\p\up}\hc_{\p\up}]\nn \\ 
    &=\int d\omega\sum_\p e^{\beta(\omega-\ep)}A_\ii (\p,-\omega)\nn \\
    & =-\frac1\pi{\rm Im}\int d\omega \sum_\p e^{-\beta \omega}G_\up^{\rm R}(\p,\omega)\nn \\
    & = -\frac1\pi{\rm Im}\int d\omega \sum_\p \frac{e^{-\beta \omega}}{\omega-\ep-\Sigma(\p,\omega)},
\end{align}
where in the last step we used the Dyson equation to relate the impurity Green's function to the self energy $\Sigma$~\cite{Fetter}. This is a useful representation since many theoretical approximations exist to calculate the self energy in various scenarios. Note that $Z_\imp$ is independent of interactions, and is thus easy to compute. Furthermore, as discussed above in Sec.~\ref{sec:specprops}, the quantity $\sum_\p  \Tr[\hr_\ri\hc^\dagger_{\p\up}\hc_{\p\up}]$ is 1 in cases where the impurity only exists in the state $\up$, in which case we simply have
\begin{align}
    e^{-\beta\Delta F}
    = -\frac{\frac1\pi{\rm Im}\int d\omega \sum_\p \frac{e^{-\beta \omega}}{\omega-\ep-\Sigma(\p,\omega)}}{\sum_\p e^{-\beta\ep}}.
\end{align}

Alternatively, we can use the sum rule in Eq.~\eqref{sum rule_b} together with the detailed balance condition in Eq.~\eqref{eq:detailed_ej_I} to obtain
\begin{align}
    e^{\beta\Delta F}=\int d\omega\, e^{-\beta\omega}I_\e(\omega).
\end{align}

\subsection{Contact}

The other thermodynamic quantity that we consider is the contact, which is the conjugate to the interaction strength~\cite{Tan2008,Braaten2008}. It appears in a variety of contexts. Notably, it governs the high-frequency tails of rf ejection spectra~\cite{Braaten2010}, the high-momentum occupation of interacting particles~\cite{Tan2008}, and it is related to the number of particles in the impurity dressing cloud~\cite{Massignan2012}. \weizhe{It has also recently been measured in the high-momentum tail of photoluminescence spectra originating from the quantum depletion of a non-equilibrium polariton condensate~\cite{Maciej2020nc}.} For further details about the contact and its relations to other thermodynamic quantities, we refer the reader to Refs.~\cite{Werner2012,Braaten2012}.

We assume that the impurity interacts with a particular medium component (e.g., identical atoms in the same hyperfine state) via a short-range contact interaction. If this is characterized by a scattering length $a$, then we define the associated impurity contact as
\begin{align}\label{eq:contact_def}
    C & \equiv \left. 8 \pi m_r \frac{\partial F}{\partial (-1/a)}\right|_{T,V} =
    \left.8 \pi m_r \frac{\partial \Delta F}{\partial (-1/a)}\right|_{T,V} ,
\end{align}
where we have used the fact that $F_0$ is independent of the interaction, and we have assumed that all chemical potentials associated with the medium are held constant. Here, we take the mass of the relevant medium particles to be $m$ and define the reduced mass $m_r\equiv mM/(m+M)$. Equation~\eqref{eq:contact_def} directly relates the contact to the impurity free energy. Since the contact can be measured in a number of ways~\cite{Tan2008,Braaten2010,Werner2012,Braaten2012}, this in principle provides an independent means of experimentally determining the impurity free energy.

We also note that
\begin{align}
    \left.\pdv{C}{T}\right|_{1/a,V} & = - 8\pi m_r \left.\pdv{S}{(-1/a)}\right|_{T,V},
\end{align}
where $S$ is the entropy. Thus we expect $\pdv{C}{T} \to 0$ in the limit of zero temperature since $S \to 0$ for any interaction.

\section{Mobile Fermi polaron}
\label{sec:Fermipolaron}

We now apply our theoretical results to a concrete example, namely an impurity interacting attractively with a Fermi gas. This is the so-called Fermi polaron, which first appeared in the context of spin-imbalanced Fermi gases~\cite{Chevy2006,Radzihovsky2010} and has since been extended to other systems such as excitons in doped two-dimensional semiconductors~\cite{Sidler2017}.

The injection spectrum of the mobile Fermi polaron contains three main features: an attractive polaron at negative energies, a repulsive polaron at positive energies~\cite{Cui2010,Massignan2011}, and a broad molecule-hole continuum in between~\cite{Kohstall2012}. While the corresponding ejection spectrum is typically harder to calculate, it is well understood in the zero-temperature limit, since the interacting many-body system is then restricted to be in its ground state. In this case, simple variational wave function approaches~\cite{Chevy2006,Combescot2007,Combescot2008,Mora2009,Punk2009,Mathy2011,Massignan2012,Nishida2015,Yi2015} have proved to be remarkably effective, being in excellent agreement with non-perturbative results from diagrammatic QMC~\cite{Prokofev2008,Vlietinck2013,Houcke2020} and fixed-node diffusion QMC~\cite{Lobo2006,Pilati2008}. In the following, we employ a finite-temperature variational approach~\cite{Liu2019} that is a generalization of the original ground-state variational ansatz in Ref.~\cite{Chevy2006}.

\subsection{Model}

To describe the Fermi polaron using the Hamiltonian~\eqref{hamiltonian} we take the medium to be fermionic. The medium-only part of the Hamiltonian takes the form
\begin{align}
    \hH_\m=\sum_\k(\ek^\m-\mu) \ha_\k^\dag \ha_\k,
\end{align}
where $\ha_\k^\dag$ creates a fermion with momentum $\k$ and mass $m$, and $\ek^\m=k^2/2m$ is the fermion dispersion. We explicitly treat the medium within the grand canonical ensemble and include the medium chemical potential $\mu$. Since we assume the medium to be in thermal equilibrium, the occupation is governed by the Fermi-Dirac distribution
\begin{align}
    n_F(\eq^\m)\equiv \Tr[\hr_\m \ha^\dag_\q\ha_\q]=\frac1{e^{\beta(\eq^\m-\mu)}+1}.
\end{align}
Therefore, the chemical potential is related to the medium density $n$ via
\begin{align}
    n=\frac1V\sum_\q n_F(\eq^\m)=-\left(\frac{mT}{2\pi}\right)^{3/2}{\rm Li}_{3/2}(-e^{\beta\mu}),
\end{align}
where ${\rm Li}$ is the polylogarithm. We also relate the density to the Fermi energy $E_F$ via $E_F=\frac{k_F^2}{2m}=\frac{(6\pi^2n)^{2/3}}{2m}$, where $k_F$ is the Fermi momentum.

In typical cold atom experiments, the short-range interactions between the medium particles and the impurity are well described by a two-channel model~\cite{Timmermans1999},
\begin{align}
\label{eq:Interaction_tc}
    \hU= \sum_\k (\ek^{\rm d} + \nu) \hd^\dagger_\k \hd_\k + \frac g{\sqrt V}\sum_{\k,\q}\left[\hd^\dagger_\k\ha_\q\hc_{\k-\q,\up} + h.c. \right],
\end{align}
in which the interaction between the (open-channel) impurity with a fermion is mediated via a so-called closed-channel molecule. Here $\hd^\dag_\k$ is the closed-channel creation operator, $\ek^{\rm d} = k^2/[2(m+M)]$ is its dispersion, $\nu$ is the bare closed-channel detuning, and $g$ is the strength of the interchannel coupling. To ensure that the model is ultraviolet convergent we introduce a momentum cutoff $\Lambda$ above which the interactions are taken to vanish. We then relate the parameters of the model to physically measurable quantities via the process of renormalization. Calculating the low-energy scattering amplitude
\begin{align}
    f(k) = -\frac1{a^{-1} + R^* k^2 +i k}
    \label{eq:scatamp}
\end{align}
within our model, we identify the $s$-wave scattering length $a$ and the range parameter $R^*$:
\begin{align}
    \frac{m_r}{2\pi a} = -\frac{\nu}{g^2} + \frac{1}{V}\sum_\k^\Lambda \frac{1}{\ek+\ek^\m}, \qquad R^* = \frac\pi{m_r^2 g^2}.
\end{align}
We remind the reader that $m_r = mM/(m+M)$. When $a>0$ there exists a single impurity-fermion bound state, with energy
\begin{align}
    \eb\equiv -\frac{\kb^2}{2m_r}=-\frac{\left(\sqrt{1+4R^*/a}-1\right)^2}{8m_r{R^*}^2}.
    \label{eq:eb}
\end{align}
This reduces to $-1/(2m_ra^2)$ in the limit where the scattering length greatly exceeds the range parameter. 

We also note that in the limit $R^*\rightarrow0$, $\hU$ reduces to
\begin{align}
   \hU_{\text{1-ch}}=\frac{u}{V}\sum_{\p,\mathbf{k},\mathbf{k}'} \ha^\dagger_{\frac{\p}{2}+\mathbf{k}} \hc^\dagger_{\frac{\p}{2}-\mathbf{k},\up} \hc_{\frac{\p}{2}-\mathbf{k}',\up} \ha_{\frac{\p}{2}+\mathbf{k}'}, 
\end{align}
where $u= -g^2/\nu$ is the effective coupling strength. In this limit, our results therefore reduce to those of the single-channel model $\hU_{\text{1-ch}}$ --- see Ref.~\cite{Liu2020_short}.

\subsection{Finite-temperature variational approach}

To model the mobile Fermi polaron, we apply the finite-temperature variational principle to impurity dynamics developed in Ref.~\cite{Liu2019} in the context of the Fermi polaron and further extended to the Bose polaron in Ref.~\cite{Field2020}. For completeness, we briefly review the method here; more details may be found in Ref.~\cite{Liu2019}. The key idea is to introduce an operator $\hbc_{\p\up} (t)$ that approximates the exact Heisenberg picture impurity operator $\hc_{\p\up} (t)$ in Eq.~\eqref{eq:heisenberg_operator}. We take $\hbc_{\p\up} (t)$ to have the form~\cite{Liu2019}
\begin{align}
    \hbc_{\p \up}(t) \simeq & \alpha_{\p;0}(t) \hc_{\p \up} + \sum_\q \alpha_{\p;\q} (t) \ha^\dagger_{\q} \hd_{\p+\q} \nn \\ &
    \hspace{5mm}+ \sum_{\k \ne \q} \alpha_{\p;\k\q}(t) \hat{f}_{\q}^\dagger \hat{f}_{\k} \,
    \hc_{\p-\k+\q,\up} .
\end{align}
$\{\alpha\}$ is a set of variational coefficients where $\alpha_{\p;0}$ is the amplitude of the bare impurity operator, $\alpha_{\p;\q}$ is the amplitude of the process where the impurity and a fermion have formed a closed-channel molecule, and $\alpha_{\p;\k\q}$ is the amplitude of the process where the impurity has scattered off a fermion. This expansion is similar in spirit to the Chevy ansatz that was introduced to extract the ground-state energy of the Fermi polaron~\cite{Chevy2006}.

In order to minimize the error arising from this approximation, we introduce the error quantity $\Delta_\p (t) = \Tr[\hr_\m \hep_\p (t) \hep_\p^\dagger (t)]$, where $\hep_\p (t) = i\partial_t \hbc_{\p\up} (t) - [\hbc_{\p\up}(t),\hH]$ is an operator that quantifies the error introduced in the Heisenberg equation of motion. The minimization condition $\partial \Delta_\p/\partial\Dot{\alpha}^*_{\p,j} = 0$ with $(j=0,\q,\k\q)$ gives~\cite{Liu2019}
\begin{align}
    &\! (E-\epsilon_\p) \alpha_{\p;0} = \frac{g}{\sqrt{V}} \sum_\q \alpha_{\p;\q} \weizhe{n_F}(\eq^\m),\nn \\
    &\! \ds (E-\epsilon_{\p;\q}) \alpha_{\p;\q} = \frac{g}{\sqrt{V}}\alpha_{\p;0} + \frac{g}{\sqrt{V}} \sum_\k \alpha_{\p;\k\q} (1-\weizhe{n_F}(\ek^\m)),\nn \\ 
    &\! \ds (E-\epsilon_{\p;\k\q}) \alpha_{\p;\k\q} = \frac{g}{\sqrt{V}} \alpha_{\p;\q},
\label{eq:kinetic_all}
\end{align}
where we have taken the stationary condition $\alpha_{\p;j}(t) = \alpha_{\p;j}(0) e^{-i E t} \equiv \alpha_{\p;j} e^{-i E t}$. We have also defined $\epsilon_{\p;\q} = \en^{\rm d}_{\p+\q} + \nu - \en^\m_\q$, $\epsilon_{\p;\k\q} = \epsilon_{\q-\k+\p} + \en^\m_{\k} - \en^\m_{\q}$.

The set of linear equations~\eqref{eq:kinetic_all} can be solved by matrix diagonalization, which generates eigenenergies $E_{\p}^{(l)}$ and eigenvectors $\{\alpha^{(l)}_{\p;j}\}.$ From these, \weizhe{we} construct stationary impurity operators $\hp_\p^{(l)} = \sum_j \alpha_{\p;j}^{(l)} \hO_j$ where $\hO_0 = \hc_{\p\up},$ $\hO_\q = \ha^\dagger_{\q} \hd_{\p+\q}$ and $\hO_{\k\q} = \ha^\dagger_{\q} \ha_{\k} \hc_{\p-\k+\q,\up}.$ We may choose these stationary operators $\hp_\p^{(l)}$ to be orthonormal, which means $\Tr[\hr_\m\hp_{\p}^{(m)} \hp_{\p}^{{(n)}\dagger}] = \delta_{m n}.$ Applying the boundary condition $\hbc_{\p\up}(0) = \hc_{\p\up}$ at $t=0$, we finally arrive at the expression for the approximate impurity operator:
\begin{align}
    \hbc_{\p\up}(t) = \sum_l \alpha_{\p;0}^{(l)*} \, \hp_\p^{(l)} e^{-i E_{\p}^{(l)} t}.\label{eq:appro_final}
\end{align}

As we shall see in the following, the spectral and thermodynamic properties of the impurity can be related to the coefficients $\alpha_{\p,0}^{(l)}$. If we knew these coefficients exactly, our results would be exact; however we will generally be calculating these by solving the set of equations~\eqref{eq:kinetic_all}.

\subsection{Spectral functions}

We first discuss the impurity spectrum calculated within the variational approach. Substituting the approximate impurity operator, Eq.~\eqref{eq:appro_final}, into Eq.~\eqref{eq:Ainj2}, we have the momentum-resolved injection spectrum~\cite{Liu2019}
\begin{align}
    A_\ii (\p,\omega) = \sum_l \abs*{\alpha_{\p;0}^{(l)}}^2\delta (\omega+\epsilon_\p - E_{\p}^{(l)}).
    \label{eq:spectrum_in}
\end{align}
Likewise, we may consider the experimentally relevant momentum-averaged spectrum
\begin{align}
    I_\ii(\omega)= \frac{\sum_{\p,l} e^{-\beta\ep} \abs*{\alpha_{\p;0}^{(l)}}^2\delta (\omega+\epsilon_\p - E_{\p}^{(l)})}{\sum_\p e^{-\beta\ep}}.
    \label{eq:spectrum_in2}
\end{align}
\weizhe{In practice, we convolve all spectral functions with a Gaussian broadening function $g(\omega) = e^{-\omega^2/(2\sigma^2)}/(\sqrt{2\pi}\sigma)$ and hence the $\delta$ function is replaced by the broadening function~\cite{Parish2016}. For example, the convoluted injection spectrum becomes
\begin{align}
    I_\ii(\omega)= \frac{\sum_{\p,l} e^{-\beta\ep} \abs*{\alpha_{\p;0}^{(l)}}^2 g (\omega+\epsilon_\p - E_{\p}^{(l)})}{\sum_\p e^{-\beta\ep}}.
    \label{eq:spectrum_in5}
\end{align}}

According to the detailed balance relations in Eq.~\eqref{eq:detailed_final}, the impurity free energy $\Delta F$ is needed to fix the overall ratio between the injection and ejection spectra. This is easiest to determine in the single-channel limit $R^*\to0$, since there is no closed-channel molecule. In this case we have $\sum_\p\Tr[\hr_\ii \hc^\dag_\p \hc_\p]=1-\sum_\p\Tr[\hr_\ii \hd^\dag_\p \hd_\p]=1$ which implies that $\int d \omega\sum_\p A_\e(\p,\omega)=1$ --- see Eq.~\eqref{sum rule_a}. We will therefore for simplicity take $R^*\to0$ throughout this section. Using the injection spectrum~\eqref{eq:spectrum_in}, the detailed balance relation~\eqref{eq:detailed_ej_F} results in
\begin{align} \label{eq:mobile_free}
    e^{-\beta\Delta F} = \frac{\sum_{\p,l} e^{-\beta E_{\p}^{(l)}} \abs*{\alpha_{\p;0}^{(l)}}^2}{\sum_\p e^{-\beta \ep}}.
\end{align}

The ejection spectrum thus takes the form
\begin{align} \label{eq:spectrum_ej}
    A_\e (\p,\omega) = \frac{\sum_l e^{-\beta E_{\p}^{(l)}} \! \abs*{\alpha_{\p;0}^{(l)}}^2\! \delta (\omega +E_{\p}^{(l)}-\epsilon_\p)}{\sum_{\p,l} e^{-\beta E_{\p}^{(l)}} \abs*{\alpha_{\p;0}^{(l)}}^2} ,
\end{align}
in the limit $R^*\to0$. Moreover, the momentum-averaged ejection spectrum is
\begin{align}
    I_\e (\omega) = \frac{\sum_{\p,l} e^{-\beta E_{\p}^{(l)}} \abs*{\alpha_{\p;0}^{(l)}}^2 \delta (\omega + E_{\p}^{(l)} - \epsilon_\p)}{\sum_{\p,l} e^{-\beta E_{\p}^{(l)}} \abs*{\alpha_{\p;0}^{(l)}}^2}.
    \label{eq:spectrum_ej2}
\end{align}
Notice the symmetry between the injection and ejection spectra: Since $\sum_l\abs*{\alpha_{\p,0}^{(l)}}^2=1$~\cite{Liu2019}, Eq.~\eqref{eq:spectrum_ej2} reduces to Eq.~\eqref{eq:spectrum_in2} if we make the exchange $\ep\leftrightarrow{E_\p^{(l)}}$.

\begin{figure}[thb]
    \centering
    \includegraphics[width=\columnwidth]{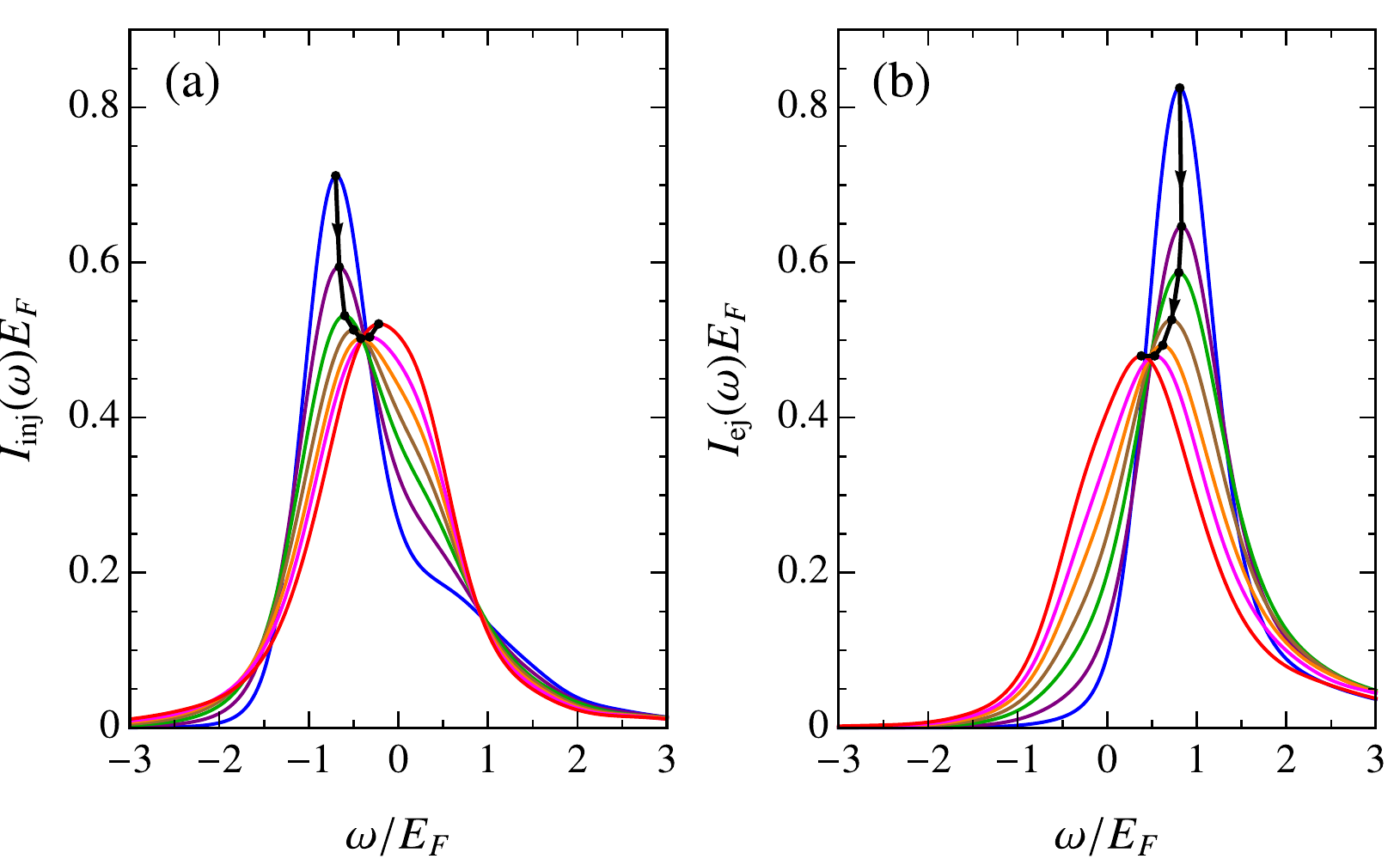}
    \caption{Injection (a) and ejection (b) spectra for the equal-mass impurity at unitarity and $R^*=0$ for various temperatures. The black dots indicate the peaks in the spectra and the \weizhe{curve} with arrows shows the tendency with increasing temperature:
    $T=0.4 T_F$, 0.6$\,T_F$, 0.8$\,T_F$, 1.0$\,T_F$, 1.25$\,T_F$, 1.5$\,T_F$ and 2.0$\,T_F.$  We have broadened the data with a Gaussian of width $\sigma_0/\veF = 0.3$.}
    \label{fig:mobile_spectra}
\end{figure}

\begin{figure}
    \centering
    \includegraphics[width=0.8\columnwidth]{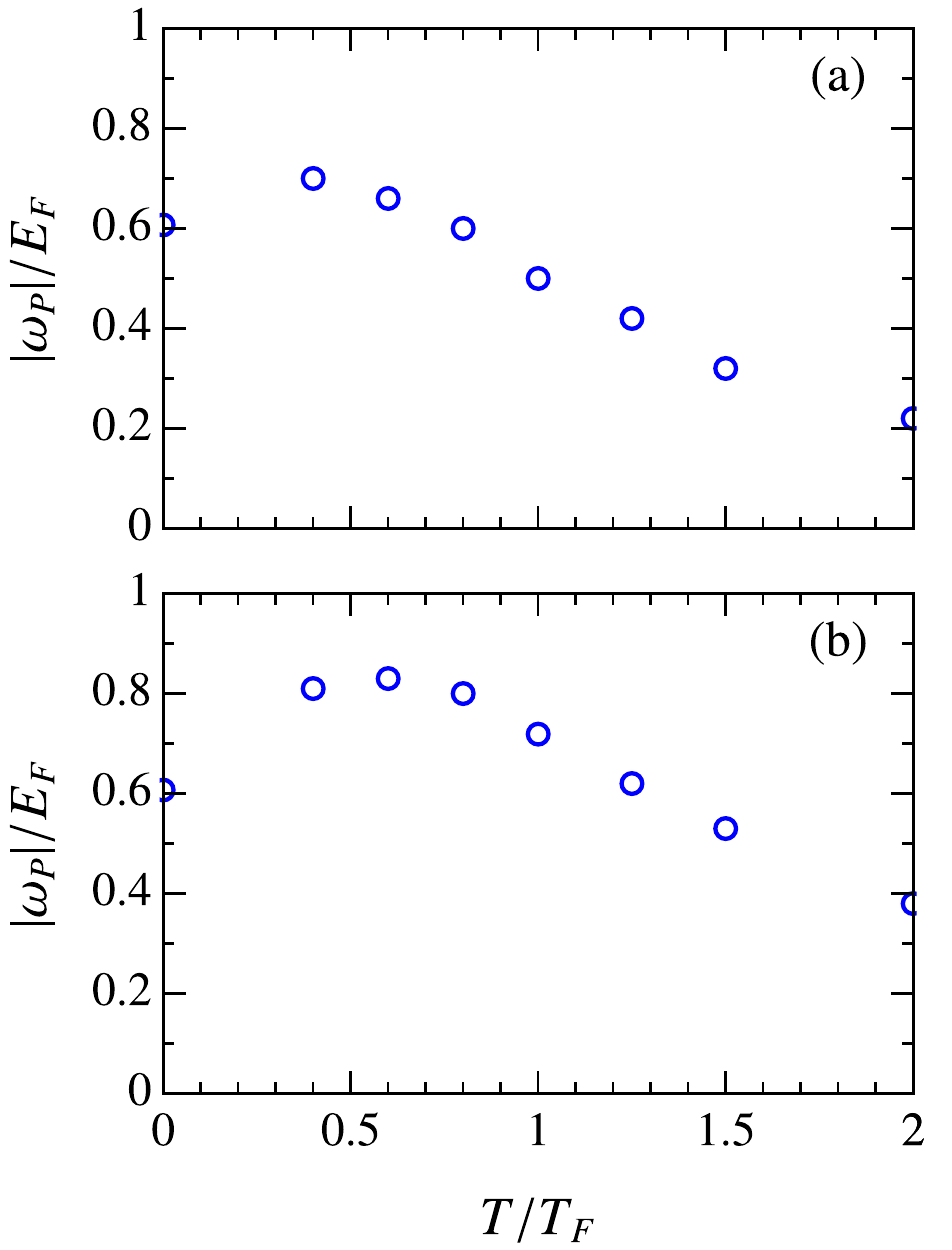}
    \caption{The peak position of the equal-mass injection (a) and ejection (b) spectra shown in Fig.~\ref{fig:mobile_spectra}.
    }
    \label{fige14}
\end{figure}

In Fig.~\ref{fig:mobile_spectra}, we show our calculated momentum-averaged injection and ejection spectra at unitarity for the equal-mass ($m=M$) Fermi polaron in the single-channel limit. Due to numerical limitations related to the potentially very large exponential prefactor in Eq.~\eqref{eq:detailed_ej_I},  we limit our attention to temperatures above $0.4T_F$ ($T_F$ is the Fermi temperature, which in our units equals the Fermi energy $E_F$) and to a relatively large Gaussian broadening of $\sigma_0/\veF = 0.3$. We see that for the lowest temperatures, the injection spectrum features an attractive polaron peak and a shoulder at positive frequency which is a combination of the molecule-hole continuum and the repulsive branch. This shoulder quickly disappears with increasing temperature, and it is also absent in the ejection spectrum due to the exponential suppression of the repulsive branch. Note that we have $I_\e(\omega) \simeq I_\ii(-\omega)$ in the high-temperature limit since the exponential prefactor in Eq.~\eqref{eq:detailed_ej_I} approaches 1 for the range of frequencies plotted in Fig.~\ref{fig:mobile_spectra}.

Figure~\ref{fige14} shows the corresponding peak positions in the spectra. At $T=0$, both injection and ejection spectra are dominated by the attractive polaron peak at $\omega=-0.61E_F$~\cite{Chevy2006} and $+0.61E_F$, respectively. We see that in both cases the peak initially shifts to a larger magnitude of frequency with increasing temperature, before eventually going to zero at high temperature. This is consistent with the results of previous finite-temperature Green's function calculations~\cite{Hu2018,Tajima2019,Mulkerin2019} at a finite density of impurities. We note, however, that the experiment in Ref.~\cite{Yan2019} observed a more dramatic change in the ejection spectrum with temperature, where the peak shifted abruptly to zero frequency at much lower temperatures than what we are finding. This is likely linked to the weak repulsive final-state interactions in experiment, which shift the relative weights of the features in the spectrum.

\subsection{Thermodynamic properties}

We now turn to the finite-temperature equation of state for the mobile impurity in the single-channel limit. This is well characterized for the ground state at zero temperature~\cite{Massignan2014}. For the equal-mass case $m=M$, the polaron undergoes a sharp single-impurity transition to a dressed dimer state at $1/(k_Fa) \simeq 0.9$~\cite{Prokofev2008,Vlietinck2013}, resulting in the loss of the polaron peak in the injection spectrum. The ejection spectrum loses the polaron peak even earlier at $1/(k_F a) \simeq 0.75$~\cite{Schirotzek2009} since the single-impurity transition is preempted by phase separation between a paired superfluid and excess fermions~\cite{Pilati2008,Mathy2011}. Note that the detailed balance condition in Eq.~\eqref{eq:detailed_bal_total} does not hold in the regime of phase separation ($0.75 \lesssim 1/(k_Fa) \lesssim 1.7$)~\cite{Pilati2008} beyond the single impurity limit, because interactions between impurities cannot be neglected in this case. However, in this section, we focus on unitarity, $1/(k_Fa) = 0$, which lies well outside this regime.

\begin{figure}[t]
    \centering
    \includegraphics[width=\columnwidth]{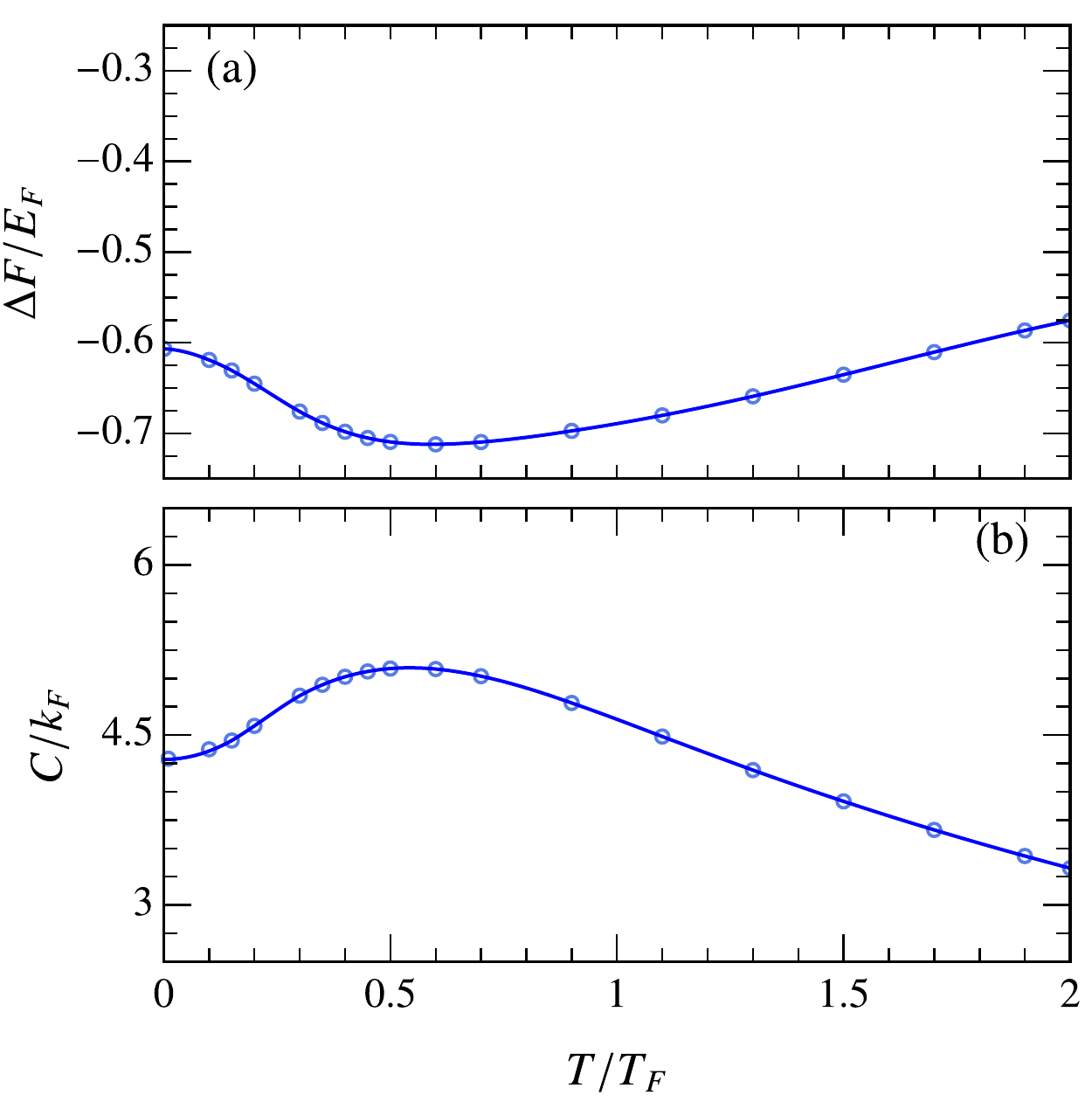}
    \caption{Temperature dependence of the equal-mass impurity free energy (a) and Tan contact (b) in the unitary regime for $R^*=0$. Our calculated data points are shown with circles, and the lines are guides to the eye. For all the data shown, the estimated relative error is $\lesssim 2\%$.}
    \label{fig:TBM_equal_contact}
\end{figure}

To obtain the free energy at arbitrary temperature, we employ Eq.~\eqref{eq:mobile_free} within the finite-temperature variational approximation. We plot in Fig.~\ref{fig:TBM_equal_contact}(a) the impurity free energy $\Delta F$ at unitarity for $m=M$ and $R^* = 0$. At zero temperature, we recover the ground-state polaron energy $-0.61 E_F$~\cite{Chevy2006}, while at finite temperature we observe a striking non-monotonic dependence on temperature. Our results suggest that the effect of the impurity-fermion attraction is strongest at $T\simeq 0.5 T_F$, before tending towards zero as $1/\sqrt{T}$ at higher temperatures.

We also compute the contact by applying the definition in Eq.~\eqref{eq:contact_def} to the free energy in Eq.~\eqref{eq:mobile_free} to obtain
\begin{align}\label{eq:TBM_contact_final}
    C = \frac{8 \pi m_r}{\beta} \frac{\partial }{\partial a^{-1}} \left[\ln \sum_{\p,l} e^{-\beta E_{\p}^{(l)}} |\alpha_{\p;0}^{(l)}|^2
    \right].
\end{align}
Equivalently, the contact can be extracted from the high-frequency tail of the ejection spectrum according to the relation~\cite{Schneider2010} 
\begin{align} \label{eq:tail}
        I_\e(\omega) \to \frac{1}{4\pi^2\sqrt{2 m_r}}\cdot\frac{C}{\omega^{3/2}} .
\end{align}
The high-frequency tail is challenging to obtain numerically 
so we use Eq.~\eqref{eq:TBM_contact_final} to calculate $C$ 
in following, but we have checked that it agrees with  Eq.~\eqref{eq:tail}.

As shown in Fig.~\ref{fig:TBM_equal_contact}(b), the contact also displays a non-monotonic dependence on temperature that mirrors that of the impurity free energy. At $T=0$, we obtain $C=4.28 k_F$, the contact for the ground-state polaron~\cite{Punk2009}, while it decreases as $1/T$ in the high-temperature limit. The non-monotonic behavior implies that the excited states of the many-body system possess a larger contact than the ground state. 
This is consistent with the existence of a molecule-hole continuum in the spectrum, which can have a larger $s$-wave pair contribution than the polaron, similar to what has been found in the three-body system~\cite{Yan2013}. Our results in Fig.~\ref{fig:TBM_equal_contact}(b) agree well with recent experiments on the unitary Fermi polaron~\cite{Yan2019} --- see Ref.~\cite{Liu2020_short} for a comparison --- and they differ from that observed in the spin-balanced case~\cite{Sagi2012,Carcy2019,Mukherjee2019}, where the contact appears to be a monotonically decreasing function of temperature. 

\section{Infinitely heavy Fermi polaron}
\label{sec:fixedimp}

We finally turn our attention to the case of an infinitely heavy Fermi polaron. This is an instructive example, since the lack of impurity recoil means that the impurity spectra and dynamics may be calculated exactly~\cite{Levitov1993jetpl,Goold2011,Knap2012,Schmidt2018}. Furthermore, at zero temperature, the system features the so-called orthogonality catastrophe, where the ground state is always orthogonal to the non-interacting state, and the spectra display power-law singularities. This means that a comparison with the variational approach, which is based on an overlap with the non-interacting state, is particularly interesting. While the $T=0$ ground state energy is known from Fumi's theorem~\cite{Fumi1955}, here we show that the impurity thermodynamic properties, i.e., the free energy and the contact, may also be calculated exactly at finite temperature.

\begin{figure*}[th]
    \centering
    \includegraphics[width=\textwidth]{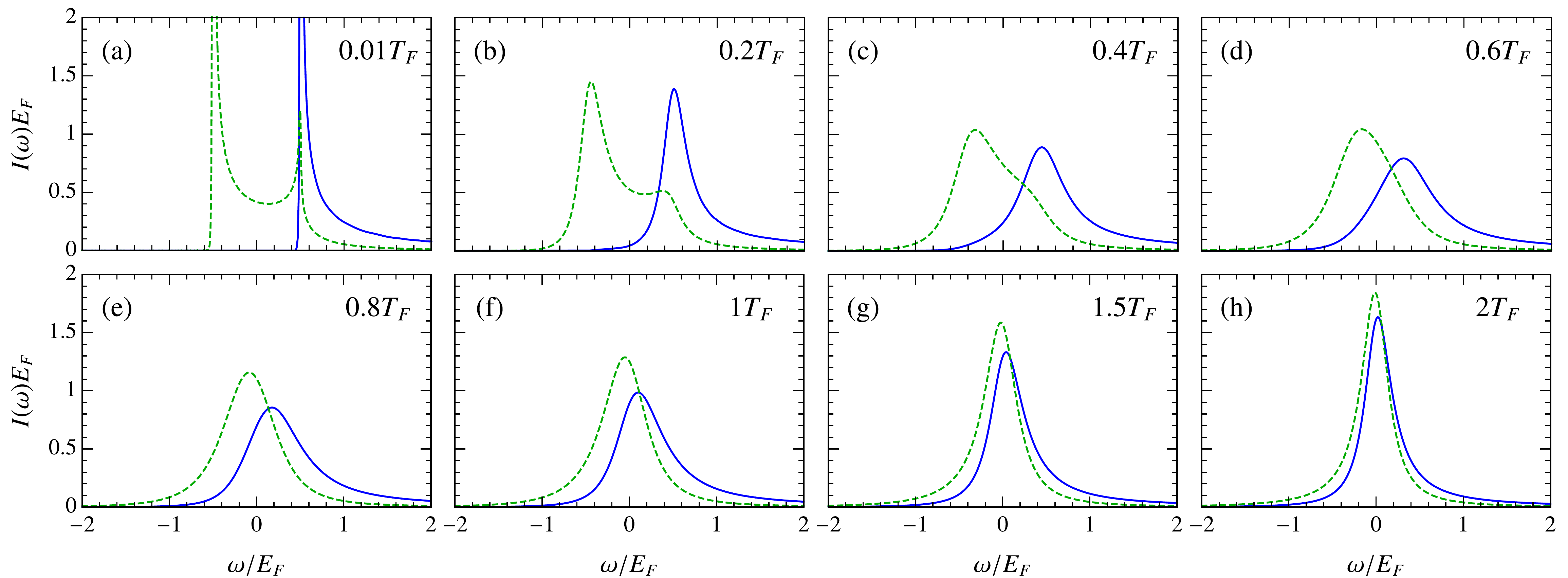}
    \caption{Exact ejection (blue solid) and injection (green dashed) spectra at various temperatures for an infinitely heavy impurity. Here we take unitarity limited interactions and the single channel limit, i.e., $1/a=R^* = 0$.}
    \label{fige6}
\end{figure*}

\subsection{Model}

In the case of an infinitely heavy impurity, we can simplify the model by noting that $\hH_\imp=0$, and that consequently we can remove the $\hc_\up$ degree of freedom. The interaction part of the Hamiltonian in Eq.~\eqref{eq:Interaction_tc} therefore reduces to~\cite{Shi2018,Yoshida2018pra}
\begin{align}
\label{eq:Interaction_inf}
    \hU= \nu \,\hd^\dagger \hd + \frac{g}{\sqrt{V}}\sum_{\q}\left[\hd^\dagger\ha_\q + \ha_\q^\dagger \hd \right],
\end{align}
where $\hd\equiv\sum_\k \hd_\k$ is a fermionic operator. Solving the two-body impurity-fermion problem now amounts to diagonalizing a bilinear Hamiltonian. The result is~\cite{Yoshida2018pra}
\begin{align}
    \hat B_\k=\sum_\p \zeta_{\k\p}\ha_\p+\eta_\k \hd,
    \label{eq:Bop}
\end{align}
where the coefficients are 
\begin{subequations}
\begin{align}
    \zeta_{\k\p}
    &= \begin{cases}
        \delta_{\k,\p} 
        + \frac{2m/V}{k^2 - p^2 + i0} T(\epsilon_\k+i0),
        & \quad \k \in \mathbb{R}^3, \\
        - \frac{1}{\sqrt{V}}\frac{2mg}{\kb^2+p^2}
        \eta_{i\kb},
        & \quad \k=i\kb,
    \end{cases}
    \label{eq:zeta} \\
    \eta_\k
    &= \begin{cases}
         T(\epsilon_\k+i0)/(g\sqrt{V}),
            & \quad \k\in \mathbb{R}^3, \\
        \sqrt{\frac{\kb R^*}{\kb R^*+1/2}},
            & \quad \k = i\kb.
    \end{cases}
    \label{eq:eta}
\end{align}
\end{subequations}
and $\kb$ is defined in Eq.~\eqref{eq:eb}. Here, $T$ is the scattering $T$ matrix, and we have chosen the boundary condition corresponding to an outgoing scattered wave, such that $T(\ek+i0) = -(2\pi/m)f(k)$, with the scattering amplitude defined in Eq.~\eqref{eq:scatamp}. We have also introduced a notation where $\k \in \mathbb{R}^3\cup\{i\kb\}$ if $a>0$ (the latter case arises from the presence of the bound state) and $\k\in\mathbb{R}^3$ if $a<0$. The sum over momentum $\p$ in Eq.~\eqref{eq:Bop} excludes the bound state contribution.

\subsection{Functional Determinant Approach}

The spectrum of an infinitely heavy impurity can be calculated exactly using a method termed the functional determinant approach (FDA), which was first introduced in Ref.~\cite{Levitov1993jetpl} and has been discussed in several recent works~\cite{Goold2011,Knap2012,Schmidt2018}. The ejection~\cite{Schmidt2018} and injection~\cite{Knap2012} spectra are
\begin{subequations}
\label{eq:FDA_basic}
\begin{align}
    I_\e (\omega) & = \int \frac{d t}{2\pi} e^{i\omega t}\de \Big[1 -\hn_\ri + \hn_\ri \,e^{i\hsh_\ri t} e^{-i\hsh_0 t} \Big], \\[2ex]
    I_\ii (\omega) & = \int \frac{d t}{2\pi} e^{i\omega t}\de \Big[1 -\hn_0 + \hn_0 \,e^{i\hsh_0 t} e^{-i\hsh_\ri t} \Big],
\end{align}
\end{subequations}
where $\hn_0 = 1/[e^{\beta (\hsh_0 -\mu)}+1]$, $\hn_\ri = 1/[e^{\beta (\hsh_\ri -\mu)}+1]$, with $\hsh_0$ and $\hsh_\ri$ the single-particle counterparts of $\hH_0$ and $\hH_\ri$, respectively. Once again, $\mu$ is the medium chemical potential. Since the FDA incorporates finite temperature and allows a separate calculation of the two spectra, it provides a platform to explicitly demonstrate the spectral relationship~\eqref{eq:detailed_final}. 

\begin{figure*}[hbt]
    \centering
    \includegraphics[width=0.95\textwidth]{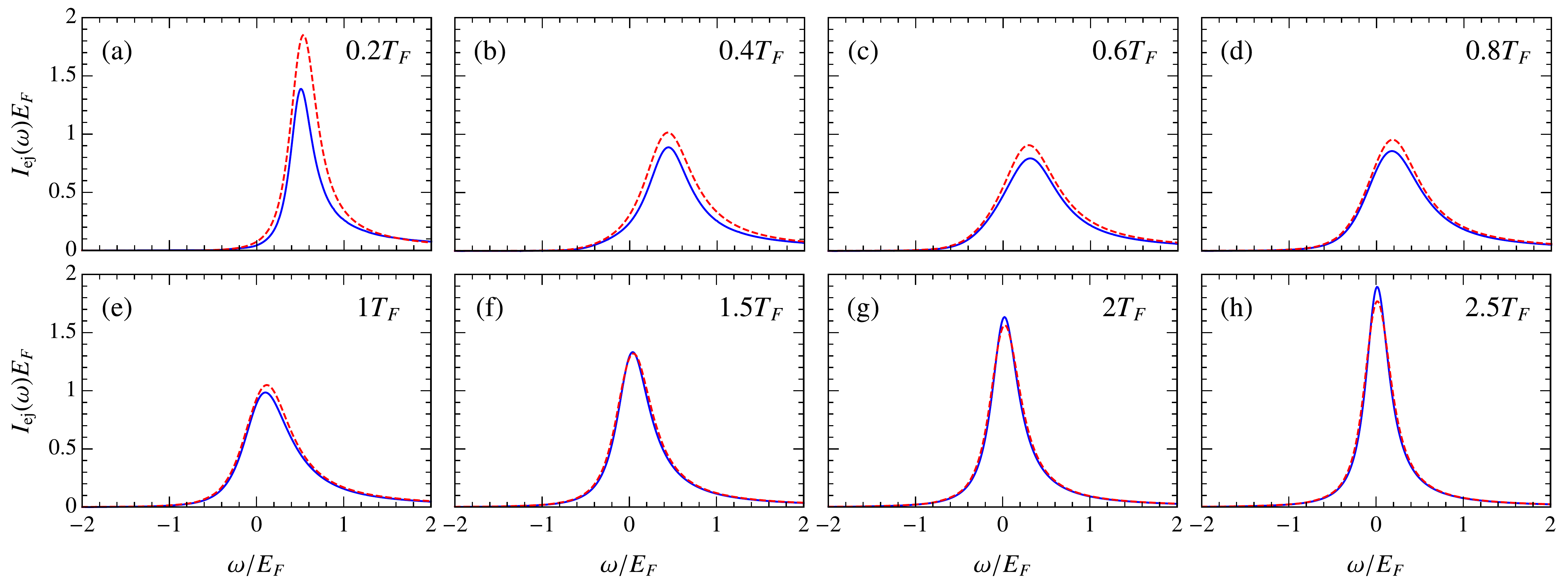}
    \caption{Comparison between the finite-temperature variational approach (red dashed) and the FDA (blue solid) ejection spectra for an infinitely heavy impurity immersed in a Fermi gas at a range of temperatures. We set $1/a = 0$ and $R^* = 0.$ For computational convenience, we applied a Gaussian broadening on all variational spectra with broadening parameter $\sigma_0 = 0.05\veF$ which is the main source of discrepancy at $T \gtrsim 2 T_F$.}
    \label{fige9}
\end{figure*}

\subsection{Spectral functions}

\begin{figure}
    \centering
    \includegraphics[width=0.8\columnwidth]{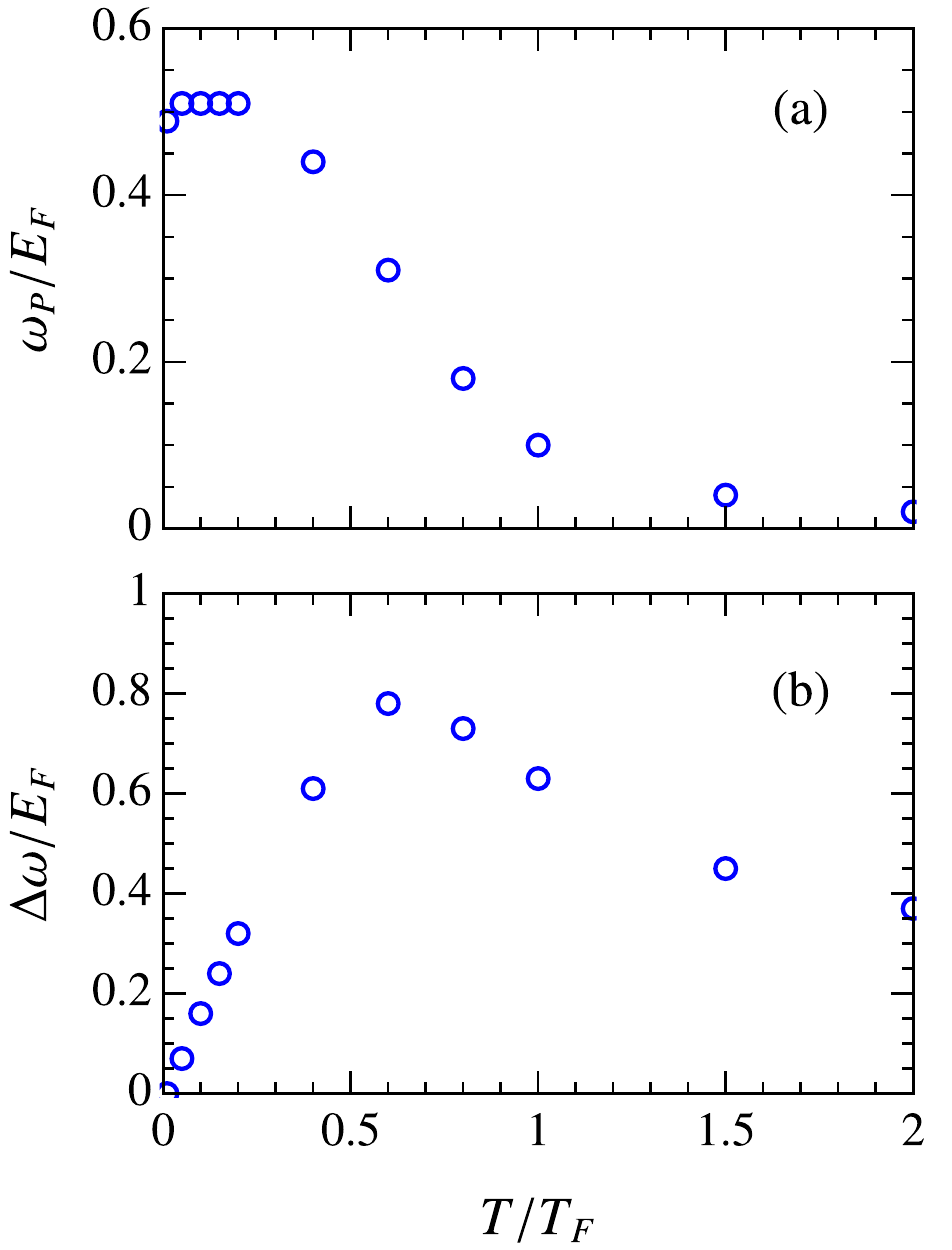}
    \caption{Temperature dependence of the ejection spectrum peak position  $\omega_{\rm P}$ (a) and full width at half maximum $\Delta \omega$ (b) for an infinitely heavy impurity. We take $1/a=0$ and $R^*=0$.}
    \label{fige13}
\end{figure}

Figure~\ref{fige6} shows the exact ejection and injection spectra at unitarity in the single-channel limit ($R^*=0$). To illustrate the evolution from a quantum to a Boltzmann gas, we show our results for a range of temperatures, from close to $T=0$ up to $T=2T_F$. At low temperatures there are significant differences between the injection and ejection spectra. In particular, the injection spectrum displays two peaks, corresponding to the attractive and repulsive branches, respectively, while the ejection spectrum contains only a single peak. This is because, from Eq.~\eqref{eq:detailed_final}, the repulsive branch in the injection spectrum is suppressed by the exponential factor $e^{-\beta\omega}$, leading to a strong suppression of spectral weight below $0.5\veF$ in the ejection spectrum. Likewise, the high-frequency tail in the ejection spectrum associated with the Tan contact~\cite{Schneider2010,Braaten2010} leads to the existence of an exponentially suppressed tail below $-0.5 \veF$ in the injection spectrum for all finite temperatures. At higher temperatures $T \gtrsim 0.6\, T_F,$ both injection and ejection spectra display a single peak that becomes increasingly narrow and centered around $\omega=0$ with increasing temperature. This conforms with the expectation that the effect of interactions vanishes in the high-temperature limit. Importantly, we have checked numerically that the injection and ejection spectra in the figure are simply related via Eq.~\eqref{eq:detailed_ej_I}, explicitly verifying the detailed balance relationship.

In Fig.~\ref{fige9}, we take advantage of the fact that we have an exact solution to benchmark the ejection spectrum calculated within our finite-temperature variational approach. We see that the low temperature $T \le 0.2 T_F$ result shows some deviation between the variational approach and FDA. However, as we increase the temperature, the agreement becomes excellent, until at high temperature, $T\gtrsim 2T_F$, there are minor differences arising due to the Gaussian broadening that we applied to accelerate our computation. Note that the corresponding comparison for the injection spectrum was carried out in Ref.~\cite{Liu2019}, where excellent agreement was demonstrated. However, the comparison for the ejection spectrum has remained inaccessible until the introduction of the spectral relationship in Eq.~\eqref{eq:detailed_final}. 

Similar to the case of the mobile Fermi polaron discussed in Sec.~\ref{sec:Fermipolaron}, we can extract the peak position in the ejection spectrum. The result is shown in Fig.~\ref{fige13}(a), where we see that initially, for $T\ll T_F$, the peak moves to slightly higher frequencies before eventually approaching zero when $T\gtrsim0.2T_F$. This is qualitatively similar to the behavior for the mobile impurity, shown in Fig.~\ref{fige14}(b), however the origin of the non-monotonicity is different. In the present case, the low-temperature peak is strongly asymmetric due to the presence of a power-law singularity in the spectrum [see Fig.~\ref{fige6}(a)] and this non-monotonicity is what translates into an initial upwards shift in peak position with temperature.

We also extract the full width at half maximum $\Delta \omega$ [see Fig.~\ref{fige13}(b)] and find a linear temperature dependence at low temperature. This is consistent with the findings of Ref.~\cite{Schmidt2018}, but it is qualitatively different from the quadratic relationship for the mobile impurity observed in Ref.~\cite{Yan2019} and theoretically found in Ref.~\cite{Mulkerin2019}. This difference stems from the lack of impurity recoil for the infinitely heavy impurity, which means that the ground state does not correspond to a well defined quasiparticle. At high temperature, $\Delta\omega\sim T^{-1/2}$ which is the same as for the mobile impurity~\cite{Yan2019}.

\subsection{Thermodynamic properties}
\label{sec:contact}

\begin{figure}[t]
    \centering
    \includegraphics[width=0.85\columnwidth]{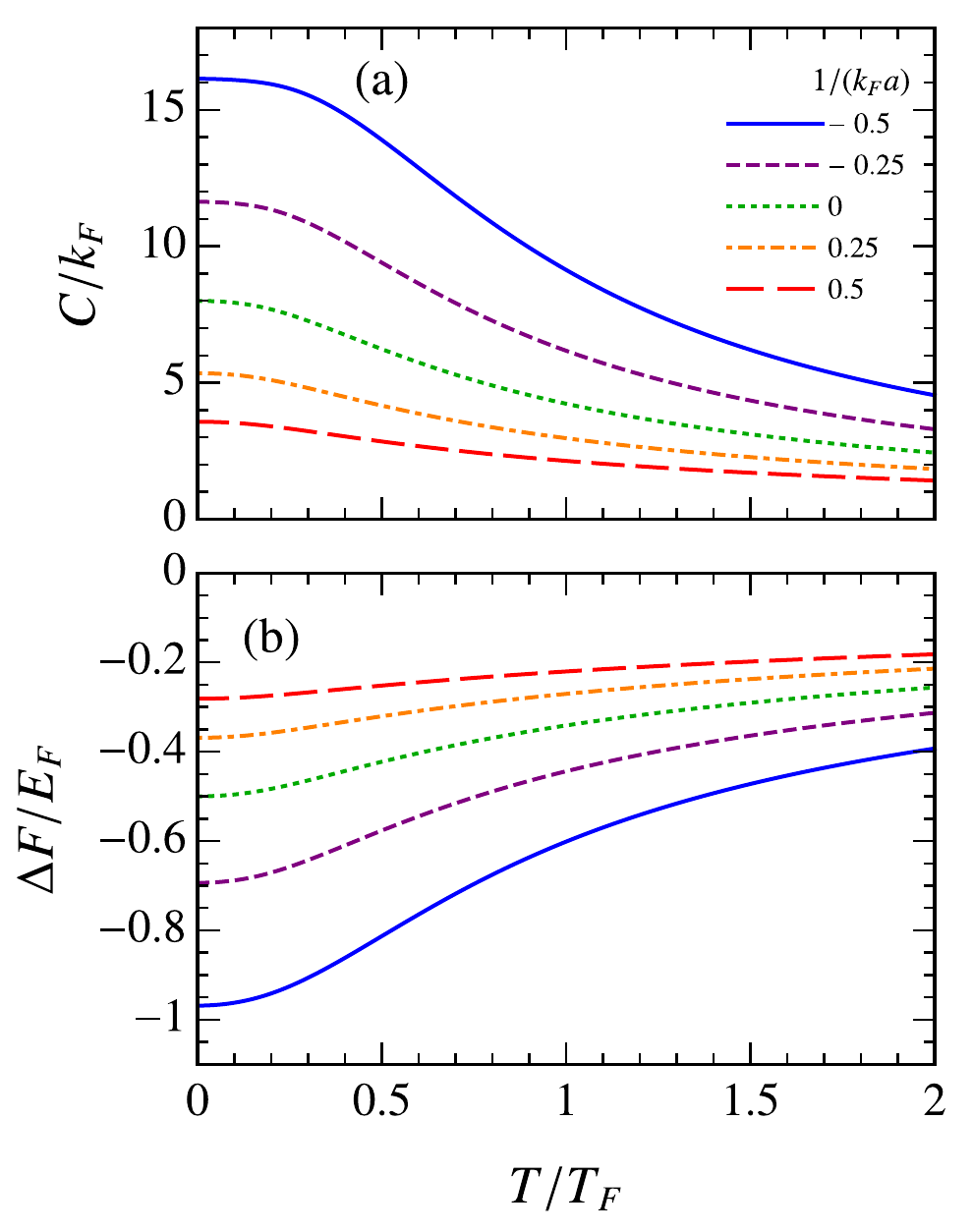}
    \caption{Temperature dependence of the contact $C$ for a fixed impurity at different scattering lengths and at $R^* =0$.}
    \label{fig:FandCinf2}
\end{figure}

\begin{figure}[t]
    \centering
    \includegraphics[width=0.85\columnwidth]{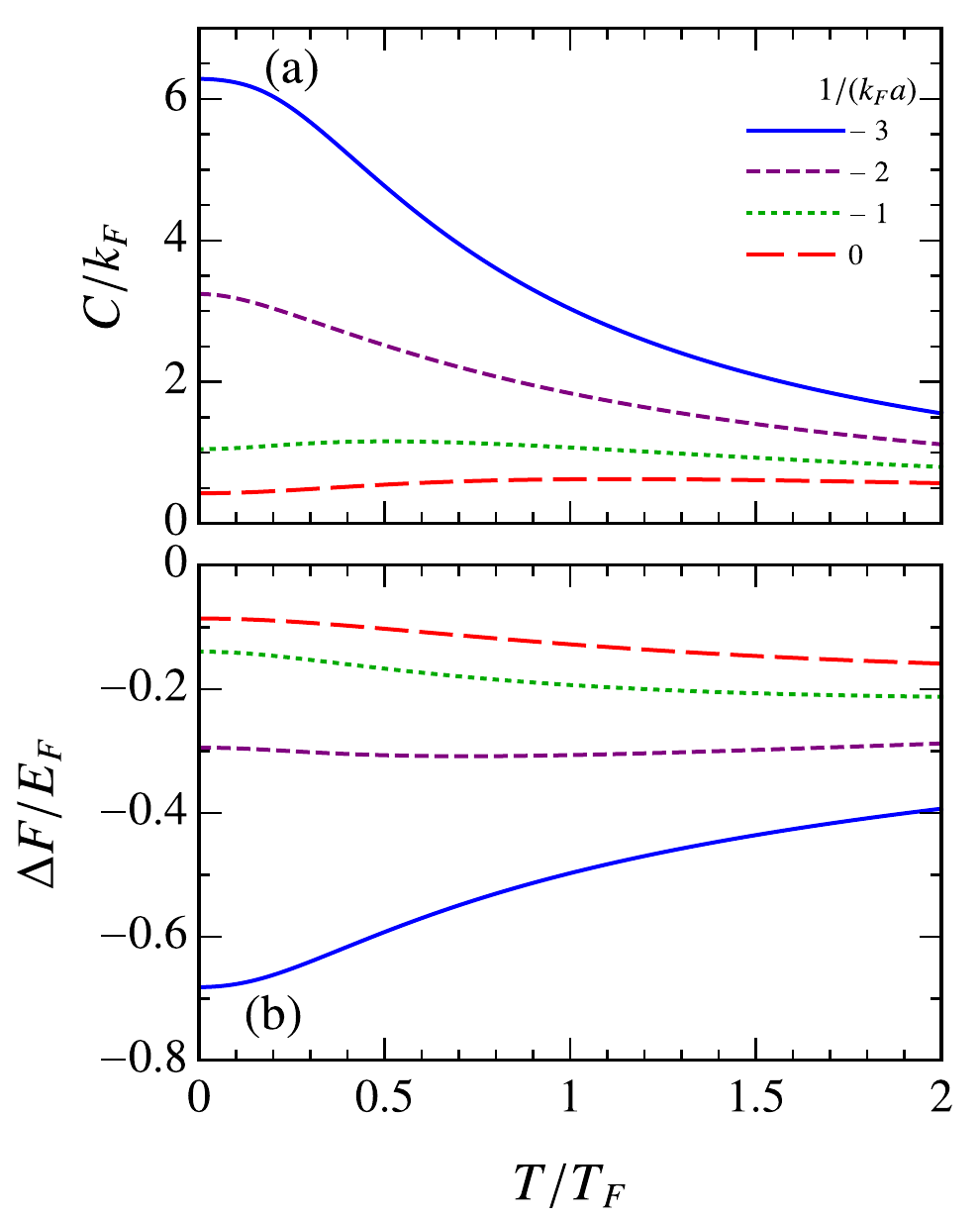}
    \caption{Temperature dependence of the contact $C$ for a fixed impurity at different scattering lengths and at $ k_FR^* = 1 $.}
    \label{fig:dFCinfR}
\end{figure}

For an infinitely heavy impurity, the interacting many-body problem becomes effectively a single-particle problem interacting with a local potential. At zero temperature, the impurity free energy is simply the ground-state energy, which can be calculated according to Fumi's theorem~\cite{Fumi1955}:
\begin{equation}
    \Delta F(T=0)= - \int_0^{E_F} dE\frac{\delta_0(E)}{\pi}  + \theta(a)\epsilon_b,
    \label{eq:Fumi}
\end{equation}
where the scattering phase shift at energy $E=\ek$ is defined as usual from $-k\cot\delta_0 (k) = 1/a+R^*k^2$.  According to Eq.~\eqref{eq:contact_def}, this results in the contact
\begin{align}
    C(T=0)&=-8\pi m\left.\pdv{\Delta F(T=0)}{(1/a)}\right|_\mu \nn \\ 
    &=8\pi m\left[\int_0^{E_F}dE\frac{\sin^2\delta_0(E)}{\pi\sqrt{2mE}}-\theta(a)\pdv{\eb}{(1/a)}\right] \nn \\
    & = \frac{16\pi^2}{V}\sum_\k^{k_F} |f(k)|^2-8\pi m\,\theta(a)\pdv{\eb}{(1/a)}.
    \label{eq:contactT0}
\end{align}

We now generalize Fumi's theorem to finite temperature. It is simplest to start from the contact
\begin{align}
    C & = 8\pi m \Tr\left[\hr_\ri \left.\pdv{\hH_\ri}{(-1/a)}\right|_{T,\mu} \right] 
    \nn \\ & 
    = 4 g^2 m^2 \Tr\left[\hr_\ri \hd^\dag\hd\right] \nn \\
    & = 4 g^2 m^2 \bigg[\sum_\k n_F(\ek^\m)
    \expval{\hd^\dagger \hd}{B_\k} \nn \\ & \hspace{16mm} +\theta(a)n_F(\eb)\expval{\hd^\dagger \hd}{B_{i\kb}}\bigg] \nn \\
    & = \frac{16\pi^2}{V}\sum_\k n_F(\ek^\m) |f(k)|^2-8\pi m\,\theta(a)n_F(\eb)\pdv{\eb}{(1/a)},
    \label{eq:Cinf}
\end{align}
where we used the diagonal form of the interacting fermions, $\ket{B_\k}\equiv \hat B^\dag_\k\ket{0}$ with $\hat B_\k$ defined in Eq.~\eqref{eq:Bop}, and $\ket{0}$ is the vacuum state. Equation~\eqref{eq:Cinf} is the natural generalization of the zero-temperature expression \eqref{eq:contactT0}.

We obtain the impurity free energy by integrating the contact in Eq.~\eqref{eq:Cinf} from the non-interacting limit:
\begin{align}
    \Delta F=-\frac1{8\pi m}\int_{-\infty}^{1/a}d(1/\tilde a)\, C.
\end{align}
Remarkably, the thermal average completely separates from the interaction part in the first term in Eq.~\eqref{eq:Cinf}. Therefore, we obtain the straightforward generalization of Fumi's theorem~\cite{Fumi1955} to finite temperature:
\begin{align}
    \Delta F=&- \int_0^\infty d E\,n_F(E) \frac{\delta_0(E)}{\pi} \nn \\ & +\theta(a)\int_{0}^{1/a}d(1/\tilde a)\, n_F(\tilde \epsilon_b) \pdv{\tilde \epsilon_b}{(1/\tilde a)}, \label{eq:Finf}
\end{align}
where the tildes indicate that the corresponding terms are evaluated at the scattering length $\tilde a$.

The expressions for the contact and impurity free energy in Eqs.~\eqref{eq:Cinf} and \eqref{eq:Finf} allow us to easily evaluate these at arbitrary interaction strength and temperature. For instance, at unitarity and in the single-channel limit ($1/a=R^*=0)$, we obtain
\begin{equation}\label{eq:contactfinal}
    C= - 4 \sqrt{2\pi m T} {\rm Li}_{\frac{1}{2}} (-e^{\beta\mu})
\end{equation}
and
\begin{align}
    \Delta F=-\frac T2\log[1+e^{\beta\mu}].
\end{align}

In Fig.~\ref{fig:FandCinf2} we show our results for the impurity free energy and the contact in the single-channel limit. We consider various values of the scattering length, including unitarity. Generically, we see that both the free energy and the contact are monotonic functions of temperature, which is qualitatively different to the equal-mass case discussed in Sec.~\ref{sec:Fermipolaron} --- see Fig.~\ref{fig:TBM_equal_contact}. This illustrates the crucial difference between the fixed and the mobile impurity: \weizhe{In the case of the mobile impurity at unitarity, there exists a well-defined quasiparticle in the spectrum (the attractive polaron) below the molecule-hole continuum~\cite{Kohstall2012}, and since the molecule-hole continuum eventually crosses the quasiparticle branch~\cite{Mora2009,Punk2009} the excited states have a larger slope with respect to $1/a$ (i.e., a higher contact), and consequently the contact initially increases with temperature. By contrast, in the present case of an infinitely heavy impurity, there is no well-defined quasiparticle ground state}, 
and therefore there is not a strong distinction between the contact of the ground state and those of the many-body continuum. 


Figure~\ref{fig:dFCinfR} illustrates that it is possible to engineer the phase shift such that the free energy and the contact become non-monotonic. Indeed, for a finite $R^*$, as considered in the figure, the absolute value of the scattering amplitude itself can become non-monotonic for negative $a$, which in turn can lead to a similar feature in the thermodynamic properties. However, note that the system still features the orthogonality catastrophe, and hence this qualitative change in the thermodynamics does not arise from a fundamental difference in the low-energy states. \weizhe{Similarly, a non-monotonic behavior has been found in a one-dimensional repulsive Fermi gas~\cite{Patu2016pra}}.

\subsection{Contact dynamics}

\begin{figure}[bht]
    \centering
    \includegraphics[width=0.85\columnwidth]{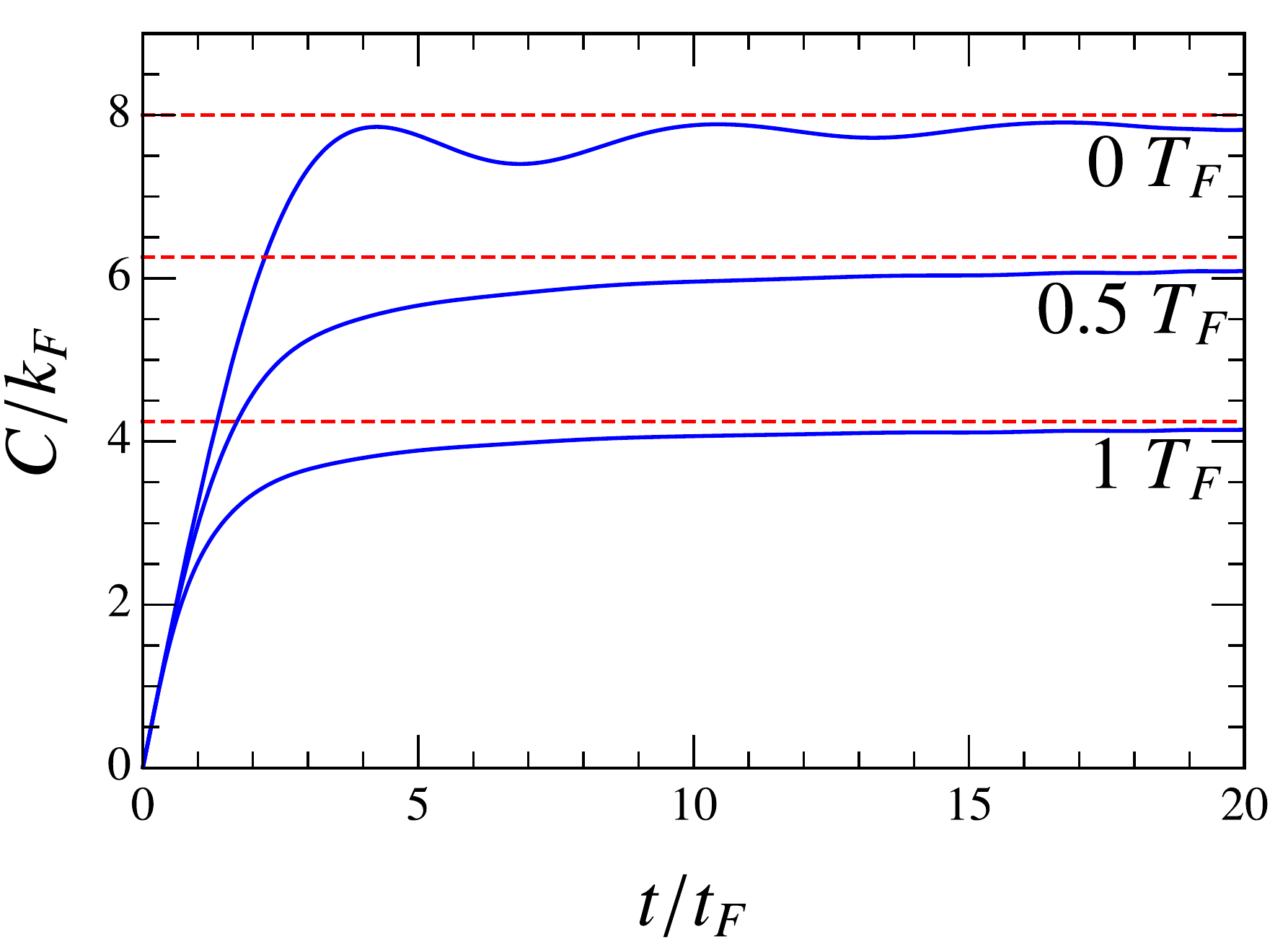}
    \caption{The exact contact dynamics (blue solid line) $C (t)$ for an infinitely heavy impurity, where we have taken $1/a = 0$ and $R^*=0$. The red dashed line is the equilibrium contact of the interacting many-body system, which is equivalent to $C(t\to\infty)$.}
    \label{fig:contact_dynamics}
\end{figure}

We end with a brief discussion of impurity dynamics when the system is out of equilibrium. In the following, we consider the ``perfect quench'' injection protocol --- i.e., the instantaneous rf transfer of the impurity into the interacting state $\up$ --- and calculate the resulting time-dependent contact. Note that this can be measured in an experiment similar to those in Refs.~\cite{Cetina2016,Fletcher2017}.

For the fixed impurity, the contact is closely related to the dimer density --- see Eq.~\eqref{eq:Cinf}. In order to apply the FDA, we need to transform $\hd^\dag \hd$ into an exponential form. This can conveniently be done using the exact transformation~\cite{Abanin2005}
\begin{align}
    \hd^\dag \hd = \frac{e^{\hd^\dag \hd} - 1}{e-1}.
\end{align}
\weizhe{This equality follows from the closed-channel dimer number operator $\hd^\dag \hd$ having only the two possible eigenvalues 0 or 1.}
%
Then, similarly to Eq.~\eqref{eq:FDA_basic}, we obtain
\begin{align}
    C(t) = \frac{\de \Big[1 -\hn_0 + \hn_0 \,e^{i\hsh_0 t} e^{\hd^\dag \hd} e^{-i\hsh_\ri t} \Big]-1}{e-1}.
\end{align}

In Fig.~\ref{fig:contact_dynamics} we show the resulting contact at unitarity as a function of time following the quench. We see that at $T=0$ the contact initially increases rapidly at the Fermi time scale $t_F\equiv 1/E_F$, until at longer times it oscillates before reaching its equilibrium value. On the other hand, the oscillations are suppressed by finite temperature. In all cases, the contact saturates to its equilibrium value at long times, in agreement with the eigenstate thermalization hypothesis~\cite{Deutsch1991pra,Srednicki1994}.

\section{Conclusion}\label{sec:concl}

To conclude, we have considered the properties of an impurity in a medium at thermal equilibrium and we have shown that the seemingly distinct processes of injection and ejection spectroscopy are intrinsically linked by a detailed balance condition. This is not obvious at first glance, since the injection spectrum often features additional peaks compared with the corresponding ejection spectrum. Our analysis has furthermore revealed that the thermodynamic properties such as the impurity free energy and contact are encoded in the ratio of the injection and ejection spectra at any frequency, rather than being linked only with the high-frequency tail of the ejection spectrum~\cite{Braaten2010} or the high-momentum occupation~\cite{Tan2008}.

For the case of the mobile Fermi polaron, we demonstrated that the impurity free energy and contact are non-monotonic functions of temperature at unitarity, with the latter being in very good agreement~\cite{Liu2020_short} with recent experiment~\cite{Yan2019}. We also obtained injection and ejection spectra at unitarity, and we found that the position of the attractive polaron peak displays a similar non-monotonic dependence on temperature.

Investigating the infinitely heavy Fermi polaron allowed us to explicitly verify the detailed balance relationship, since the injection and ejection spectra can be calculated exactly in an  independent manner. We also compared the exact ejection spectra with our finite-temperature variational approach~\cite{Liu2019}, and obtained good agreement, further illustrating the utility of this technique in investigating a variety of impurity problems. We furthermore demonstrated that the impurity free energy corresponds to the generalization of Fumi's theorem to finite temperature, and used this to obtain the exact impurity contact at any temperature and interaction strength. As opposed to the case of the mobile Fermi polaron, the impurity free energy and contact are monotonic functions of temperature at unitarity, which we have argued is connected to 
the presence of the orthogonality catastrophe in this problem.

The detailed balance condition linking ejection and injection spectroscopy and their relation to the impurity free energy are independent of dimensionality and of the details of the medium. \weizhe{It can also be used to link spectroscopic protocols in other systems such as for excitonic impurities in charge-doped atomically thin semiconductors~\cite{Sidler2017}, where absorption and photoluminescence corresponds, respectively, to injection and ejection spectroscopy.} The theory that we have presented in this work therefore provides a versatile tool for the investigation of the numerous quantum impurity scenarios emerging in the cold-atom context and beyond.

\section*{Acknowledgements}

We are grateful to Haydn Adlong and Martin Zwierlein for insightful discussions. J.~L., Z.~Y.~S., and M.~M.~P. acknowledge support from the Australian Research Council via Discovery Project No. DP160102739. W.~E.~L., J.~L. and M.~M.~P. acknowledge support from the Australian Research Council Centre of Excellence in Future Low-Energy Electronics Technologies (Grant No. CE170100039). J.~L. is supported through the Australian Research Council Future Fellowship No. FT160100244.

\appendix

\section{Detailed balance condition}
\label{app:Aejinj_def}

Here we provide the details of how to arrive at the detailed balance condition, Eq.~\eqref{eq:detailed_balance}, from the definition of the ejection and injection spectral functions in Eq.~\eqref{eq:fermigrule}. Starting from the ejection spectral function, we have
\begin{align}
    A_\e(\p,\omega)&
    =\sum_{n,\nu}\frac{e^{-\beta E_{\nu}}}{Z_\ri}|\!\bra{n}\hc_{\p\up}\ket{\nu}\!|^2 \delta(E_{\nu n\p}+\omega)\nn \\
    & =\sum_{n,\nu}\frac{e^{-\beta (E_n+\ep-\omega)}}{Z_\ri}|\!\bra{n}\hc_{\p\up}\ket{\nu}\!|^2 \delta(E_{\nu n\p}+\omega)\nn \\
    & =e^{\beta (\omega-\ep)}\sum_{n,\nu}\frac{e^{-\beta E_n}}{Z_\ri}|\!\bra{n}\hc_{\p\up}\ket{\nu}\!|^2 \delta(E_{\nu n\p}+\omega)\nn \\
    & =\frac{Z_\m Z_\imp}{Z_\ri}e^{\beta\omega} n_{\rm B}(\p)A_\ii(\p,-\omega)
\end{align}
where in the last step we used the definition of the injection spectral function, as well as the impurity Boltzmann distribution density in Eq.~\eqref{eq:nB}.

\section{Detailed balance condition for interacting initial and final states}
\label{app:detailed_int}

In the main text, we have assumed that we have \weizhe{a single impurity that can exist in two states}, $\ket{\up}$ and $\ket{\down}$, where only state $\ket{\up}$ interacts with the medium. We now show that we can derive a detailed balance condition for the case when state $\ket{\down}$ also interacts with the medium. Note that, in this case, the detailed balance condition is not momentum resolved.

In analogy with the formalism in Section~\ref{sec:impspectra}, we define interacting impurity+medium states for impurity in state $\ket{\sigma}$ as $\ket{\nu_\sigma}$ with associate energy $E_{\nu_\sigma}$ and partition function $Z_{\ri,\sigma}$. The transfer rate from state $\ket{\up}$ to $\ket{\down}$ at a given frequency $\omega$ is now
\begin{align}
    {\cal I}_{\up\to\down}(\omega)=&2\pi\OR^2\sum_{\nu_\up,\nu_\down,\p}\frac{e^{-\beta E_{\nu_\up}}}{Z_{\ri,\up}}|\!\bra{\nu_\down}\hat c^\dag_{\p\down}\hat c_{\p\up}\ket{\nu_\up}\!|^2\nn \\ &\hspace{12mm}\times \delta(\omega+E_{\nu_\up}-E_{\nu_\down}).
\end{align}
Likewise, we have
\begin{align}
    {\cal I}_{\down\to\up}(\omega)=&2\pi\OR^2\sum_{\nu_\down,\nu_\up,\p}\frac{e^{-\beta E_{\nu_\down}}}{Z_{\ri,\down}}|\!\bra{\nu_\up}\hat c^\dag_{\p\up}\hat c_{\p\down}\ket{\nu_\down}\!|^2\nn \\ &\hspace{12mm}\times \delta(\omega+E_{\nu_\down}-E_{\nu_\up}).
\end{align}
Therefore, we conclude that
\begin{align}
{\cal I}_{\up\to\down}(\omega)= \frac{Z_{\rm int,\down}}{Z_{\rm int,\up}}e^{\beta\omega}{\cal I}_{\down\to\up}(-\omega).
\end{align}
This reduces to Eq.~\eqref{eq:detailed_bal_total} when state $\ket{\down}$ does not interact with the medium.

\section{Relating spectral functions to the time-dependent impurity operator}
\label{app:spectral}

Here we show the details of how to obtain Eq.~\eqref{spectral_functions_all} from Eq.~\eqref{eq:fermigrule}. Take $A_\e (\p,\omega)$ as an example:
\begin{align}
    A_\e (\p,\omega)\equiv & \sum_{n,\nu} \frac{e^{-\beta E_\nu}}{Z_{\rm int}} |\!\bra{n} \hc_{\p\up} \ket{\nu}\!|^2 \delta (E_\nu - E_{n} - \ep + \omega).
    \label{eq:Aeapp}
\end{align}
First, we write the delta function as an integral
\begin{align}
    \delta (E_\nu - E_{n} - \ep + \omega)=\Re\int_0^\infty \frac{dt}\pi\,
    e^{i(E_\nu - E_{n} - \ep + \omega)t}.
\end{align}
Then we use 
\begin{align}
    &\sum_n\frac{e^{-\beta E_\nu}}{Z_{\rm int}} \bra{\nu}\hc_{\p\up}^\dag\!\ketbra{n} \hc_{\p\up} \ket{\nu}e^{i(E_\nu - E_{n})t} \nn \\
    = & \sum_n\bra{\nu}\hr_\ri e^{i\hH_\ri t}\hc_{\p\up}^\dag e^{-i\hH_0 t}\ketbra{n}\hc_{\p\up}\ket{\nu}\nn \\
    = &\bra{\nu}\hr_\ri \hc_{\p\up}^\dag(t)\hc_{\p\up}(0)\ket{\nu},
    \label{eq:appstep}
\end{align}
where in the last step we replaced $H_0$ by $H_\ri$ when it acts on the medium-only states, removed the complete set of medium states, and used the definition of the time-dependent impurity operator in Eq.~\eqref{eq:heisenberg_operator}. Gathering terms in Eqs.~\eqref{eq:Aeapp}-\eqref{eq:appstep}, we arrive at Eq.~\eqref{eq:Aej2}, namely
\begin{align}
    A_\e(\p,\omega)& = \Re\int_0^\infty \frac{dt}\pi e^{i(\omega-\epsilon_\p) t}\Tr\left[\hr_\ri\hc_{\p\up}^\dagger(t)\hc_{\p\up}(0) \right].
\end{align}
In a completely analogous fashion we arrive at Eq.~\eqref{eq:Ainj2} for the injection spectral function.

\section{Spectral functions for a finite density of impurities}
\label{app:spectral_GC}

In this appendix, we discuss the decomposition of the grand canonical impurity spectral function into a particle and a hole contribution, as discussed in Sec.~\ref{sec:finitedens}, as well as the detailed balance condition between these. We start by inserting the definition of the spin-$\up$ Green's function, Eq.~\eqref{eq:Gfinitedens}, into the equation for the total spectral function:
\begin{align}
    {\cal A}(\p,& \omega) = -\Im[{\cal G}^{\rm R}_\up(\p,\omega)]/\pi\nn \\
     &=\Re\int_0^\infty \frac{dt}\pi\, e^{i(\omega-\mu_\up)t}\Tr[\hr_{\rm G} \{ e^{i\hK t}\hc_{\p\up} e^{-i\hK t},\hc^\dagger_{\p\up}\}\weizhe{_{\pm}}] \nn \\
     & = \Re\int_0^\infty \frac{dt}\pi\,e^{i(\omega+E_\xi-E_\nu) t}\nn \\ & \hspace{10mm}\times\sum_{\nu,\xi}\abs{\bra{\xi}\!c_{\p\up}\!\ket{\nu}}^2  \left(\expval{\hr_{\rm G}}{\xi}\weizhe{\pm}\expval{\hr_{\rm G}}{\nu}\right),
\end{align}
where the trace is taken over eigenstates of $\hat K$, and we inserted an additional complete set of states of $\hat K$. The states denoted $\ket{\nu}$ obviously contain one more impurity particle than the states denoted $\ket{\xi}$, which means that the dependence on chemical potential in the phase cancels. Since $\Re\int_0^\infty \frac{dt}\pi\,e^{i(\omega+E_\xi-E_\nu) t}=\delta (\omega + E_\xi - E_\nu )$, we obtain Eq.~\eqref{eq:spectral_GC}:
%
\begin{align}
    {\cal A}(\p,\omega) & = \overbrace{\sum_{\xi,\nu} \abs{\bra{\xi} \hc_{\p\up} \ket{\nu}}^2 \delta (\omega + E_\xi - E_\nu ) \expval{\hr_{\rm G}}{\xi}}^{{\cal A}_+ (\p,\omega)} \nn \\ 
    & \weizhe{\pm} \underbrace{\sum_{\xi,\nu} \abs{\bra{\xi} \hc_{\p\up} \ket{\nu}}^2 \delta (\omega + E_\xi - E_\nu ) \expval{\hr_{\rm G}}{\nu}}_{{\cal A}_- (\p,\omega)}.
\end{align}
%

The detailed balance condition for a finite impurity density can be found by using the properties of the delta functions, together with the expressions for the expectation values of the density matrix:
\begin{align}
    &\delta (\omega + E_\xi - E_\nu ) \expval{\hr_{\rm G}}{\xi} \nn \\
    & =\delta (\omega + E_\xi - E_\nu ) e^{-\beta(E_\xi-\mu_\up N_{\up,\xi})} \nn \\ 
    & = \delta (\omega + E_\xi - E_\nu ) e^{-\beta(E_\nu-\omega-\mu_\up (N_{\up,\nu}-1))} \nn \\
    & = e^{\beta(\omega-\mu_\up)}\delta (\omega + E_\xi - E_\nu ) \expval{\hr_{\rm G}}{\nu}.
\end{align}
Thus, we have the detailed balance condition in Eq.~\eqref{eq:detailed_GC}, namely ${\cal A}_-(\p,\omega)=e^{-\beta (\omega-\mu_\up)}{\cal A}_+(\p,\omega)$.


\begin{thebibliography}{87}
\bibitem [{foo()}]{foot:volume}%
  \BibitemOpen
  \bibinfo {note} {Note that $n_B(\p)$ can be thought of as a density since
  $Z_\imp=\sum_p e^{\beta \ep}\sim V$ implies that
  $n_B(\p)\sim1/V$.}\BibitemShut {Stop}%
\end{thebibliography}
\end{document}